\documentstyle[aps,preprint,tighten,epsfig]{revtex}
\begin{document}
\draft


\title{
Production of $\bbox{\omega}$ and $\bbox{\phi}$ Mesons in Near-Threshold
$\bbox{\pi\,N}$ Reactions:
Baryon Resonances and Validity of the OZI Rule
}


\author{
A.I.~Titov$^{a,b,}$\footnote{
E-mail address: {\tt atitov@thsun1.jinr.ru}},
B.~K\"ampfer$^{a,}$\footnote{
Corresponding author.\\
{} \hspace*{2mm} E-mail address: 
{\tt kaempfer@fz-rossendorf.de} (B. K\"ampfer).},
B.L. Reznik$^{c,}$\footnote{
E-mail address: {\tt reznik@dvgu.ru}}
}

\address{
$^a$ Forschungszentrum Rossendorf, PF 510119, 01314 Dresden,
Germany\\
$^b$ Bogolyubov Laboratory of Theoretical Physics, JINR,
Dubna 141980, Russia
\\
$^c$ Far-Eastern State University, Sukhanova  9, Vladivostok 690090,
Russia
}

\maketitle

\begin{abstract}
Results of a combined analysis are presented for
the production of $\omega$ and $\phi$ mesons
in $\pi N$  reactions in the near-threshold region
using throughoutly a conventional ''non-strange''
dynamics  based on such processes  which are allowed
by the non-ideal $\omega-\phi$ mixing.
We show that strong interferences of the $t$ (meson exchange)
and $s$ and $u$ (nucleon and nucleon resonance) channels
differ significantly in $\omega$ and $\phi$ production amplitudes.
This leads to a decrease of the relative yields
in comparison with expectations
based on one-channel models with standard $\omega - \phi$ mixing.
We find a strong and non-trivial difference between
observables in  $\omega$ and $\phi$ production
reactions  caused by the different role of the nucleon and
nucleon resonance amplitudes.
A series of predictions
for the experimental study of this effect is presented.\\[3mm]
{\it PACS:} 13.75.-n; 14.20.-c; 21.45.+v\\
{\it keywords:} hadron reactions; phi and omega production; threshold
behavior.
\end{abstract}

\section{Introduction}

The present interest in a combined study of the $\omega$ and
$\phi$ meson production in different
elementary reactions is mainly related to the
investigation of the hidden strangeness degrees of freedom in the nucleon.
Since the $\phi$ meson is thought to consist mainly of strange
quarks, its production
should be suppressed according to the OZI rule \cite{OZI}
if the entrance channel
does not possess a considerable admixture of strangeness.
The standard OZI rule violation is described by the deviation
from the ideal $\omega-\phi$ mixing by the angle
$\Delta\theta_V\simeq 3.7^0$ \cite{PDG98}, which is a measure of
the small contribution of light $u,\bar u$ and $d,\bar d$ quarks
in the $\phi$ meson, or strange $s,\bar s$ quarks in the $\omega$
meson.
Thus, the ratio of $\omega$ to $\phi$ production
cross sections is expected to be
$R^2_{\omega/\phi}\simeq {\rm ctg}^2\Delta\theta_V\simeq
2.4\times10^{2}$.

Indeed, the recent experiments on the proton annihilation at rest
(cf.\ \cite{Ellis} for references and a compilation of data) point to
a large apparent violation of the OZI rule,
which is interpreted \cite{Ellis,Ellis99} as a
hint to an intrinsic $s\bar s$ component in the proton.
However, the data can be
explained as well by modified meson exchange models \cite{LZL93}
without introducing any strangeness component in the nucleon or
OZI rule violation mechanisms.

On the other hand,
the analysis of the $\pi N$ sigma term \cite{sigmaterm} suggests that
the proton might contain a strange quark admixture as large as 20\%.
Thus this issue remains  controversial.
Therefore it is tempting to look for other
observables \cite{Ellis,newexp,TOY97_98} that are sensitive to the
strangeness content of the nucleon. Most of them are related to
a possible strong interference of delicate $s \bar s$
knock-out (or shake-off)
amplitudes and the ``non-strange'' amplitude which is caused by OZI rule
allowed processes, or by  processes wherein the standard OZI rule
violation comes from the $\phi - \omega$ mixing.

A detailed analysis of the current status of the OZI rule in
$\pi N$ and $NN$ reactions has been presented recently in \cite{SC00}.
It is shown in \cite{SC00} that existing data for the $\omega$ and
$\phi$ meson production in $\pi N$ reactions give for the ratio of
averaged amplitudes
the value of $R_{\omega/\phi} = 8.7\pm1.8$,
which is much smaller than the standard
OZI rule violation value of $R^{OZI}_{\omega/\phi}=15.43$
and may be interpreted
as a hint to non-zero strangeness components in the nucleon.

Obviously, reliable information on a manifestation of
hidden strangeness in the combined study of $\phi$ and $\omega$ production
processes can be obtained only when the conventional, i.e. non-exotic,
amplitudes have been understood quantitatively.
The reaction $\pi N \to V N$ with $V = \omega$, $\phi$ has the evident
advantage to be a simple hadronic reaction
representing a subprocess, e.g., in $N N \to V N N$ reactions.
The study of the former reactions is one of the
objectives of the present work.
The dominant conventional processes
are depicted in Fig.~1, where
(a) is the $t$ channel meson exchange process,
while (b) depicts the $s$, $u$ nucleon and nucleon resonance channels.
When taking separately each of the amplitudes, the ratio of $\omega$
to $\phi$ production amplitudes is proportional to
${\rm ctg }\Delta\theta_V=15.43$
and the question arises how their coherent sum
can result in a deviation from this value.

Note that most of the previous considerations of the possible violation of
the OZI rule in hadronic reactions are based on one-channel models
in the spirit of \cite{Lipkin76,Faelt_Wilkin}.
But if one assumes that the process is a coherent sum of
at least two amplitudes, say for example, a meson exchange and
a nucleon term, then the result may be different from the naive
expectation.
Indeed, let us suppose for a moment that,
because of some hadronic
dynamics, the nucleon term 
for the $\phi$ production is suppressed
relative to the meson exchange term.
Then the ratio of $\omega$ to $\phi$ amplitudes becomes
\begin{equation}
R_{\omega / \phi}
= {\rm ctg} \Delta \theta_V \,
\frac{\vert 1 + R^\omega_{N/M} \vert}{\vert 1 + R^\phi_{N/M} \vert}
\approx  {\rm ctg} \Delta \theta_V \,
\vert  1 +   R^\omega_{N/M} \vert ,
\end{equation}
where  $R^{\omega , \phi}_{N/M}$
are the ratios of nucleon and meson exchange amplitudes.
Thus, one can get immediately an enhancement (suppression) of
$R_{\omega / \phi}$, as compared to the OZI rule prediction,
for a constructive
(destructive) interference between the two $\omega$ amplitudes
and for $|R^\phi_{N/M}| \ll |R^\omega_{N/M}|<1 $.
Our previous study \cite{TKR00} of $\phi$ production shows that the
interference between meson exchange and nucleon terms is destructive.
Assuming the same for $\omega$ production,
$R_{\omega/\phi}$ must decrease,
even without any speculation on a
strangeness content in the nucleon.
In case of considering the $\omega$ production, however,
this two-channel model is by no means longer adequate,
because one has to include various strong resonance
channels. The ratio $R^\omega_{(N+N^*)/M}$ becomes complex which
may change the above estimates in any direction.

The resonance contribution to vector meson production
has its own interest because it might affect significantly
in-medium polarization operators and the corresponding dilepton
emissivity of hadronic matter \cite{FP97,LWF99,PM00}.
Therefore its
detailed study in elementary $\pi N$ processes is another objective
of the present work. An important step in this direction has been done
recently by Riska and Brown in \cite{RB00}, where the
relevant $\pi N N^*$, $\omega N N^*$ and $\phi N N^*$
coupling constants are expressed in terms of the corresponding
couplings to nucleons using 
a quark model.
Our study here exploits essentially the findings of
\cite{RB00}. The role of
the low-lying nucleon resonances in a combined study of $\rho^0$ and
$\omega$ production
based on the relativistic coupled-channel model has been studied in
Refs.~\cite{LWF99,SLF00}.
Some aspects of the $\rho$ meson spectral function
in the nucleon resonance model have been discussed in
\cite{Mosel_Post_Leupold}.
The contribution of the higher resonances to the $\omega$
photo-production has been analyzed in Ref.~\cite{OTL00}.

\begin{figure}
\centering
{\epsfig{file=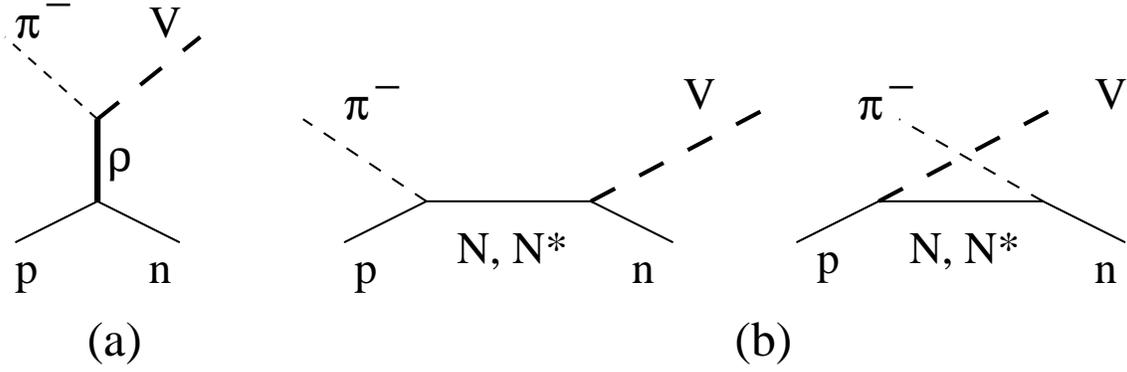, width=15cm}}
\bigskip
\caption{
Diagrammatic representation of the $\pi^- p\to V n$ reaction
mechanisms with $V = \omega$, $\phi$.
(a) meson exchange diagram with
vector meson emission from the $V\rho\pi$ vertex,
(b) nucleon and nucleon-resonance vector-meson production in the
$V NN$ and  $V NN^*$ vertices.}
\label{fig:1}
\end{figure}

Our analysis of the reaction $\pi N \to V N$ is based on calculations
of the diagrams in Fig.~\ref{fig:1}.
While the diagrams in Fig.~1 look like usual Feynman diagrams it
should be stressed that they give a guidance of how to obtain from an
interaction Lagrangian of hadronic  fields a covariant parameterization
of observables in strict tree level approximation. Additional ingredients
are needed to achieve an accurate description of data within such a framework.
In particular, the vertices needs to be dressed by form factors.
The early theoretical studies~\cite{KO,TKS99}
show, indeed, that predictions for hadronic observables are very sensitive
to the parameters of the form factors
which can not be fixed unambiguously without adjustments
relying on the corresponding experimental data.

In Refs.~\cite{Nakayama98,Nakayama99} the parameters of
$V NN$ interactions have been determined by analyzing commonly
the reactions $pp\to pp \omega$  and $pp\to pp \phi$
and the corresponding new DISTO data \cite{DISTO}
at a given beam energy assuming the same production mechanism
without resonance contributions.
In a previous paper \cite{TKR00}, in order to constrain the parameter
space further, we performed a combined analysis
of the reactions $\pi^- p\to n\phi$ and $pp\to pp\phi$
at the same energy excess.
A good description of available data has been achieved.
We did not consider in \cite{TKR00} the $\omega$ production
for which the production mechanism
is more complicated because of the resonance contributions
shown in Fig.~1b. Therefore,
the problem of the validity of the OZI rule was beyond
the scope of considerations in \cite{TKR00}.

In this paper we attempt a different approach with
the goal to check the validity of the OZI rule in a combined study
of the related reactions  $\pi N\to N\omega$ and $\pi N\to N\phi$
using the known data within the same interval of excess energies
$10 \cdots 100$ MeV and taking into account the
nucleon resonance channels.

Our paper is organized as follows.
In Section II, we define the effective Lagrangians, derive expressions for
the amplitudes of the processes shown in Fig.~1
and discuss the parameter fixing.
In Section III the results of numerical
calculations and predictions are presented.
The summary is given in Section IV.


\section{Amplitudes}

The differential cross section of the reaction
$\pi^-p\to V n $ with $V=\omega , \phi$ (cf. Fig.~1)
has the obvious form in standard notation
\begin{eqnarray}
\frac{d\sigma}{d\,\Omega}
=\frac{1}
{64\pi^2s}\,\frac{|{\bf q}|}{|{\bf k}|}{|T|^2},
\label{CS_pi}
\end{eqnarray}
where 
$k = (E_\pi,\, {\bf k})$
and $q=(E_V, {\bf q})$
are the four-momenta of the pion and the
vector meson in the center of mass system (c.m.s.);
the squared invariant amplitude
${|T|^2}$ includes the average and sum over
the initial and final spin states, respectively.
We denote the four-momenta of the initial (target) and final (recoil)
nucleons by $p$ and $p'$;
$\Omega$ and $\theta$ are
the solid and polar angles of the produced vector meson in the c.m.s.;
$s = (p+k)^2$ is the usual Mandelstam variable.

We also consider the spin density matrix $\rho_{rr'}$ which defines the
angular distribution in the decays
$\omega, \phi \to e^+e^-$,  
$\omega \to \pi^+\pi^-\pi^0$ and 
$\phi \to K^+K^-$.
It has a simple form in
the system where the vector meson is at rest (for details see
\cite{TKR00}). The decay angles $\Theta$, $\Phi$ are defined as polar
and azimuthal angles of the direction of the three-momentum of one of
the decay particles in the vector meson's rest frame. For the
$\omega\to \pi^+\pi^-\pi^0$ decay, $\Theta$ is the polar angle
of the direction of the vector product
$[{\bbox k}_{\pi^+}\times{\bbox k}_{\pi^-}]$, where
${\bbox k}_{\pi^+}$ and ${\bbox k}_{\pi^-}$ are the momenta of  $\pi^+$ and
$\pi^-$ mesons, respectively~\cite{ABBHHMC68}.
The $e^+ e^-$ decay distribution
integrated over the azimuthal angle $\Phi$, ${\cal W}(\cos\Theta)$,
depends only on the diagonal matrix elements $\rho_{00}$, $\rho_{11} =
\rho_{-1-1}$, normalized as $\rho_{00} + 2\rho_{11} = 1$, according to
\begin{eqnarray}
{\cal W}^{e^+e^-}(\cos\Theta) & = &
\frac34
\left[ 1 + \rho_{00} +(1-3\rho_{00}) \cos^2 \Theta \right].
\label{W}
\end{eqnarray}
The corresponding distributions for the hadronic decays
$\phi \to K^+K^-$, $\omega\to \pi^+\pi^-\pi^0$
are
\begin{eqnarray}
{\cal W}^{h}(\cos \Theta) = \frac32
\left[ 1- \rho_{00} - (1 - 3\rho_{00}) \cos^2 \Theta \right].
\label{Kkd}
\end{eqnarray}
In our calculation we choose the quantization axis ${\bf z}$ along
the beam momentum.

\subsection{Effective Lagrangians}

In calculating the invariant amplitudes for the basic processes shown
in Fig.~1 we use the following effective interaction Lagrangians.\\
(i) interactions in the meson exchange process (Fig.~1a):
\begin{eqnarray}
{\cal L}_{V \rho\pi} & = &
g_{V \rho \pi} \, \epsilon^{\mu\nu\alpha\beta} \,
\partial_\mu^{(V)}
V_\nu \, {\rm Tr} ( \partial_\alpha \rho_\beta \pi), \\
{\cal L}_{\rho NN} & = &
-\,\,{g_{\rho NN}}\bar\psi_{N}
\left(\gamma_\mu - \frac{\kappa_{\rho NN}}{2M_N}
\sigma_{\mu\nu} \partial^\nu_{(\rho)}
\right)\rho^\mu\,\psi_{N},
\label{L_meson}
\end{eqnarray}
where  ${\rm Tr}(\rho \pi)
= \rho^0 \pi^0 + \rho^+ \pi^- + \rho^- \pi^+$,
and $\pi$ and $\rho^\mu$ denote the pion and rho meson fields.
The partial derivatives $\partial_\mu^{(V)}$ and $\partial^\nu_{(\rho)}$
are meant to act only on the corresponding fields $V^\mu$ and
$\rho^\mu$; $\epsilon^{\mu\nu\alpha\beta}$ is the Levi-Civita symbol.\\
(ii) interactions in the baryonic channels (Fig.~1b):
\begin{eqnarray}
{\cal L}_{MNN}^{N_{\frac12^+}(940)\, N}
& = &
\bar\psi_N\, \left[ - \frac{f_{\pi NN}}{m_\pi}
\gamma_5\gamma_\mu\,\partial^\mu
\bbox{\pi \cdot \tau}\,\,
-\,\,{g_{V NN}}
\left(\gamma_\mu - \frac{\kappa_{V NN}}{2M_N}
\sigma_{\mu\nu}\partial^\nu
\right)V^\mu\,\right]\psi_{N},
\label{N940}\\
{\cal L}_{MNN^*}^{N_{\frac12^+}(1440)\, P_{11}}
& = &
\bar\psi_N\, \left[ -\frac{f^{1440}_{\pi NN^*}}{m_\pi}
\gamma_5\gamma_\mu\,\partial^\mu
\bbox{\pi \cdot \tau}\,
-\,\,{g^{1440}_{V NN^*}}
\,(\gamma_\mu + \partial_\mu \not\hskip-0.7mm\!{\partial} \,m_V^{-2} )
V^\mu\,\right]\psi_{N^*} \,\, +  \,{\rm h.c.},
\label{N1440}\\
{\cal L}_{MNN^*}^{N_{\frac32^-}(1520)\, D_{13}}
& = &
\bar\psi_N \, \left[
i \frac{f^{1520}_{\pi NN^*}}{m_\pi}
\,\gamma_5\,\partial^\alpha
\bbox{\pi \cdot \tau}\,\,
+\,\, \frac{g^{1520}_{V NN^*}}{m_V^2}
\sigma_{\mu\nu}\partial^\nu\partial^\alpha
\,V^\mu\,\right]{\psi_{N^*}}_\alpha
\,\, + \,{\rm h.c.},
\label{N1520}\\
{\cal L}_{MNN^*}^{N_{\frac12^-}(1535)\,S_{11}}
& = &
\bar\psi_N\, \left[
-\frac{f^{1535}_{\pi NN^*}}{m_\pi}
\gamma_\mu\,\partial^\mu
\bbox{\pi \cdot \tau}\,\,
-\,\,{g^{1535}_{V NN^*}}
\,\gamma_5
\,(\gamma_\mu + \partial_\mu\not\hskip-0.7mm\!{\partial}\, m_V^{-2} )
V^\mu\,\right]\psi_{N^*}
\,\, + \,{\rm h.c.},
\label{N1535}\\
{\cal L}_{MNN^*}^{N_{\frac12^-}(1650)\, S_{11}}
& = &
\bar\psi_N\, \left[
-\frac{f^{1650}_{\pi NN^*}}{m_\pi}
\,\gamma_\mu\,\partial^\mu
\bbox{\pi \cdot \tau}\,\,
-\,\,g^{1650}_{V NN^*}\gamma_5
\,(\gamma_\mu + \partial_\mu\not\hskip-0.7mm\!{\partial}\, m_V^{-2} )
V^\mu\,\right]\psi_{N^*}
\,\, + \,{\rm h.c.},
\label{N1650}\\
{\cal L}_{MNN^*}^{N_{\frac52^-}(1675)\,D_{15}}
& = &
\bar\psi_N\, \left[
-\frac{f^{1675}_{\pi NN^*}}{m_\pi^2}
\,\partial^\alpha\partial^\beta
\bbox{\pi \cdot \tau}\,\,
+\,\,\frac{g^{1675}_{V NN^*}}{m_V^2}
\,\epsilon^{\alpha \gamma \mu \nu}\,\gamma_\nu \,\partial_\gamma
\partial^\beta
\,V_\mu\,\right]{\psi_{N^*}}_{\alpha\beta}
\,\, + \,{\rm h.c.},
\label{N1675}\\
{\cal L}_{MNN^*}^{N_{\frac52^+}(1680) F_{15}}
& = &
\bar\psi_N\, \left[
-i\frac{f^{1680}_{\pi NN^*}}{m_\pi^2}
\,\gamma_5
\partial^\alpha\partial^\beta
\bbox{\pi \cdot \tau}\right.\,\,\nonumber\\
&& \left. \quad\quad
+\,\,\frac{g^{1680}_{V NN^*}}{m_V^2}
\,(\gamma_\mu + \partial_\mu\not\hskip-0.7mm\!{\partial}\, m_V^{-2} )
\,\partial^\alpha\partial^\beta V^\mu\,
\right]{\psi_{N^*}}_{\alpha\beta}\,\, + \,{\rm h.c.},
\label{N1680}\\
{\cal L}_{MNN^*}^{N_{\frac32^-}(1700) D_{13}}
& = &
\bar\psi_N\, \left[
i\frac{f^{1700}_{\pi NN^*}}{m_\pi}
\,\gamma_5
\partial^\alpha
\bbox{\pi \cdot \tau}\,\,
+\,\,\frac{g^{1700}_{V NN^*}}{m_V^2}\,
\sigma_{\mu\nu}\partial^\nu\partial^\alpha
\,V^\mu\,\right]{\psi_{N^*}}_\alpha
\,\, + \,{\rm h.c.},
\label{N1700}\\
{\cal L}_{MNN^*}^{N_{\frac32^-}(1720) P_{13}}
& = &
\bar\psi_N \, \left[
i\frac{f^{1720}_{\pi NN^*}}{m_\pi}
\,\partial^\alpha
\bbox{\pi \cdot \tau}\,\,\right.
\nonumber\\
& & \left. \quad\quad - \frac{g^{1720}_{V NN^*}}{M_{N^*}+M_N}
\,\gamma_5 \, \left( \gamma_\mu\partial^\alpha
- g_\mu^\alpha\,\not\hskip-0.7mm\!{\partial}
\right)V^\mu\,
\right]{\psi_{N^*}}_{\alpha}\,\, + \,{\rm h.c.},
\label{N1720}
\end{eqnarray}
where ${\bbox \pi}$, $V_\mu$, $\psi_N$ and $\psi_{N^*}$ are
the pion iso-vector,
iso-scalar vector meson $V=\omega, \, \phi$,
nucleon and Rarita-Schwinger
nucleon resonances field operators, respectively,
and the subscript $M$ stands for ''meson''.
$\bbox{\tau}$ denotes the Pauli matrix.
Note that these interaction Lagrangians do not contain
partial derivatives of the fields $\psi_{\cdots}$ and their
adjoints $\bar \psi_{\cdots}$. The notation of the masses
is self-explaining.
We use the convention of Bjorken and Drell \cite{BD}
in definitions of $\gamma$ matrices and the
spin matrix $\sigma_{\mu \nu}$.
The expressions Eqs.~(\ref{N940} - \ref{N1720})
are based on \cite{RB00}.\footnote{
Our notation differs from that in \cite{RB00}
by the substitutions
$\partial\to i\partial$ and
$ig_{VNN}\to - g_{VNN}$
keeping the relative phases between $f_{\pi NN}$ and $f_{\pi NN^*}$
and  $g_{V NN}$ and $g_{V NN^*}$ the same as in \cite{RB00}.
We also
express ${\cal L}_{MNN^*}$ in a manifestly gauge-invariant form.}
As in \cite{RB00} we include here all **** resonances up to 1720 MeV
according to \cite{PDG98} and the *** resonance
$N_{\frac 32 ^-}(1700)D_{13}$ as well. The contribution of the
*** resonance $N_{\frac 12 ^-}(1710)P_{13}$ will be discussed below.

All coupling constants with off-shell mesons are dressed by monopole
form factors \cite{Bonn}
$F_i=(\Lambda_i^2-m_i^2)/(\Lambda_i^2-k_i^2)$,
where $k_i$ is the four-momentum of the exchanged meson.
Following the scheme of the meson photo-production in \cite{Hab97}
we assume that the $V NN$ and $VNN^*$ vertices
must be dressed by form factors for off-shell baryons
\begin{eqnarray}
F_{B}(r^2) =
\frac{\Lambda_B^4}{\Lambda_B^4 + (r^2-M^2_B)^2 },
\label{cutN}
\end{eqnarray}
where $M_B$ is the baryon mass and $r$ is
the four-momentum  of the virtual baryons $B = N, N^*$
in Fig.~1b.

\subsection{Invariant amplitudes}

The total invariant amplitude is sum
of the meson exchange, nucleon and nucleon resonance channels,
\begin{eqnarray}
T_{\lambda}=
T^{(M)}_{\lambda} + T^{(N)}_{\lambda} + T^{(N^*)}_{\lambda},
\end{eqnarray}
where $\lambda = 0, \pm 1$ is the polarization projection of the
produced vector meson.
The amplitude for the meson exchange channel in Fig.~1a reads
\begin{eqnarray}
T^{(M)}_{\lambda} = K^{\pi N} \,
\epsilon^{\alpha \beta \gamma \delta}
\left[ \bar u(p')\, \Gamma^{(\rho)}_\delta (k_{(\rho)}) u(p) \right] \,
q_\alpha k_\gamma \, \varepsilon_\beta^{* \, \lambda} \, I_\pi ,
\label{T_pi_M}
\end{eqnarray}
where
\begin{eqnarray}
\Gamma^{(\rho)}_\alpha (k_{(\rho)})
& = &
\gamma_\alpha
 + i\frac{\kappa_{\rho NN}}{2M_N} \sigma_{\alpha \beta} \,
k_{( \rho )}^{\beta},
\label{Gamma} \\
K^{\pi N} (k_{(\rho)})
& = &
- \frac{g_{\rho NN}\,g_{V \rho\pi} }{k_{(\rho)}^2-m_\rho^2}
\frac{\Lambda_{\rho NN}^2-m_\rho^2}{\Lambda_{\rho NN}^2-k_{(\rho)}^2}
\frac{{\Lambda^{\rho \, 2}_{V\rho\pi}}-m_\rho^2}
{{\Lambda^{\rho 2}_{V\rho\pi}}-k_{(\rho)}^2}
\end{eqnarray}
with
$k_{(\rho)} = p' - p$ as the virtual $\rho$ meson's four-momentum;
$\varepsilon_\beta^\lambda$ is the vector meson's polarization four-vector,
$I_\pi$ denotes the isospin factor being
$\sqrt{2}$ (1) for a $\pi^-$
($\pi^0$) meson in the entrance channel;
the nucleon spin indices are not displayed;
$\alpha, \beta, \gamma, \delta, \mu, \nu, \tau$ are Lorentz indices
throughout the paper (not to be mixed with the notations of meson species
$\pi, \rho, \omega, \phi$),
and $u(p)$ denotes bispinors
(not to be mixed with the Mandelstam variable $u$).

The invariant amplitudes for the nucleon and resonant channels
in Fig.~1b have the following form
\begin{eqnarray}
T^{(N)}_{ \, \lambda} & = & g_{V NN}\,\frac{f_{\pi NN}}{m_\pi}\,
\bar{u}(p')\,{\cal A}^\mu(N)\,
{u}(p) \, \varepsilon^{* \, \lambda}_\mu \, I_\pi, \\
T^{(N^*)}_{ \, \lambda} & = & g_{V NN^*}\,\frac{f_{\pi NN^*}}{m_\pi}\,
\bar{u}(p')\,{\cal A}^\mu({N^*})\,
{u}(p) \, \varepsilon^{* \, \lambda}_\mu \, I_\pi,
\label{T_pi_N*}
\end{eqnarray}
where the operators ${\cal A}_\mu(N)$ and ${\cal A}_\mu({N^*})$
follow from the effective Lagrangians
of Eqs.~(\ref{N940} - \ref{N1720}) as
\begin{eqnarray}
{\cal A}_\mu(N^{940} \, )
&=&
i\frac{\Gamma^V_\mu (-q)\Lambda({p}_L,M_{N^*})
\gamma_5\not\hskip-0.7mm\!{k}\,
F_N(s)}{s-m_N^2}
+
i\frac{\gamma_5\not\hskip-0.7mm\!{k}\, \Lambda({p}_R,M_{N^*})
\Gamma^V_\mu (-q) F_N(u)}{u-m_N^2}, \label{Ampl1}\\ \nonumber\\
{\cal A}_\mu({N^{1440}}) &=&
i\frac{\gamma_\mu\Lambda({p}_L,M_{N^*})
\gamma_5\not\hskip-0.7mm\!{k}\,
F_{N^*}(s)}{s-M_{N^*}^2 +i\Gamma_{N^*}M_{N^*}}
+
i\frac{\gamma_5\not\hskip-0.7mm\!{k}\,\Lambda({p}_R,M_{N^*})
\gamma_\mu F_{N^*}(u)}{u-M_{N^*}^2 + i\Gamma_{N^*}M_{N^*}},
\label{AmplN1440}\\ \nonumber\\
{\cal A}_\mu({N^{1520}})
& = &
-\frac{\sigma_{\mu\nu}q^\nu q^\alpha
\Lambda_{\alpha\beta}({p}_L,M_{N^*})\gamma_5 k^\beta
F_{N^*}(s)}{m_V^2(s-M_{N^*}^2 +i\Gamma_{N^*}M_{N^*})}\nonumber\\
&~& \hspace*{4cm}
- \frac{\gamma_5 k^\alpha\Lambda_{\alpha\beta}({p}_R,M_{N^*})
\sigma_{\mu\nu}\,q^{\nu}q^{\beta}
F_{N^*}(u)}{m_V^2(u-M_{N^*}^2 + i\Gamma_{N^*}M_{N^*})},\\  \nonumber\\
{\cal A}_\mu({N^{1535}}) &=&
i\frac{\gamma_5\gamma_\mu\Lambda({p}_L,M_{N^*})
\not\hskip-0.7mm\!{k}\,
F_{N^*}(s)}{s-M_{N^*}^2 +i\Gamma_{N^*}M_{N^*}}
+
i\frac{\not\hskip-0.7mm\!{k}\,\Lambda({p}_R,M_{N^*})
\gamma_5\gamma_\mu F_{N^*}(u)}{u-M_{N^*}^2 + i\Gamma_{N^*}M_{N^*}},\\ \nonumber\\
{\cal A}_\mu({N^{1650}})
& = &
i\frac{\gamma_5\gamma_\mu\Lambda({p}_L,M_{N^*})
\not\hskip-0.7mm\!{k}\,
F_{N^*}(s)}{s-M_{N^*}^2 +i\Gamma_{N^*}M_{N^*}}
+
i\frac{\not\hskip-0.7mm\!{k}\,\Lambda({p}_R,M_{N^*})
\gamma_5\gamma_\mu F_{N^*}(u)}{u-M_{N^*}^2 + i\Gamma_{N^*}M_{N^*}},\\    \nonumber\\
{\cal A}_\mu({N^{1675}})
& = &
-\frac{\epsilon^\alpha_{\tau \mu \nu} q^\tau q^\beta k^\gamma k^\delta }
{m_\pi m_V^2}
\left(
\frac{\gamma^\nu
\Lambda_{\alpha \beta, \gamma \delta}({p}_L,M_{N^*})
F_{N^*}(s)}{s-M_{N^*}^2 +i\Gamma_{N^*}M_{N^*}}\right.\nonumber\\
&~& \hspace*{4cm}
+ \left.
\frac{\Lambda_{\gamma \delta, \alpha \beta}({p}_R,M_{N^*}) \gamma^\nu
F_{N^*}(u)}{u-M_{N^*}^2 + i\Gamma_{N^*}M_{N^*}}
\right),\\  \nonumber\\
{\cal A}_\mu({N^{1680}})
& = &
-i \frac{q^\alpha q^\beta k^\gamma k^\delta }
{m_\pi m_V^2}
\left(
\frac{\gamma_\mu
\Lambda_{\alpha \beta, \gamma \delta}({p}_L,M_{N^*})\,\gamma_5
F_{N^*}(s)}{s-M_{N^*}^2 +i\Gamma_{N^*}M_{N^*}}\right.\nonumber\\
&~& \hspace*{4cm}
+ \left.
\frac{\gamma_5\Lambda_{\gamma \delta, \alpha \beta}({p}_R,M_{N^*})\,
F_{N^*}(u)}{u-M_{N^*}^2 + i\Gamma_{N^*}M_{N^*}}
\right),\\ \nonumber\\
{\cal A}_\mu({N^{1700}})
& = &
-\frac{\sigma_{\mu\nu}q^\nu q^\alpha
\Lambda_{\alpha\beta}({p}_L,M_{N^*})\,\gamma_5 k^\beta
F_{N^*}(s)}{m_V^2(s-M_{N^*}^2 +i\Gamma_{N^*}M_{N^*})}\nonumber\\
&~& \hspace*{4cm}
- \frac{\gamma_5 k^\alpha\Lambda_{\alpha\beta}({p}_R,M_{N^*})
\sigma_{\mu\nu} q^{\nu}q^{\beta}
F_{N^*}(u)}{m_V^2(u-M_{N^*}^2 + i\Gamma_{N^*}M_{N^*})},\\   \nonumber\\
{\cal A}_\mu({N^{1720}})
& = &
i\frac{\gamma_5 \,(q^\alpha \gamma_\mu - g_\mu^\alpha\not\hskip-0.7mm\!{q})\,
\Lambda_{\alpha\beta}({p}_L,M_{N^*})k^\beta
F_{N^*}(s)}{(M_{N^*}+M_N)(s-M_{N^*}^2 +i\Gamma_{N^*}M_{N^*})},
\nonumber\\
&~& \hspace*{4cm}
+ i \frac{k^\beta \Lambda_{\beta\alpha}({p}_R,M_{N^*})
\gamma_5 \, (q^\alpha \gamma_\mu - g_\mu^\alpha\not\hskip-0.7mm\!{q})\,
F_{N^*}(u)}{(M_{N^*}+M_N)(u-M_{N^*}^2 + i\Gamma_{N^*}M_{N^*})},
\label{AmplA}
\end{eqnarray}
with $p_L=p+k$, $p_R=p-q$
 and $\Gamma^V_\mu$ as in Eq.~(\ref{Gamma}) but with
$\kappa_{VNN}$.

The resonance propagators in Eqs.~(\ref{AmplN1440} - \ref{AmplA})
are defined by the conventional  method \cite{IZ}
assuming the validity of the spectral decomposition
\begin{eqnarray}
\psi_{N^*}(x)=\int \frac{d^3{\bf p} }{(2\pi)^3\sqrt{2E_p}}
\left[
a_{{\bf p},r} \, u_{N^*}^r(p)e^{-ipx} +
b^{+}_{{\bf p},r} \, v_{N^*}^r(p)e^{+ipx}
\right].
\end{eqnarray}
The finite decay width $\Gamma_{N^*}$ is introduced into
the propagator denominators by substituting
$M_{N^*} \to M_{N^*} + \frac12 \Gamma_{N^*}$.
Therefore, the operators $\Lambda({p},M)$ are defined as
\begin{eqnarray}
\Lambda({p},M)
& = &
\frac12 \sum_r \left(
(1+\frac{p_0}{E_0})u^r({\bf p},E_0)\otimes\bar u^r({\bf p},E_0)
\right. \nonumber \\
& &  \left. \hspace*{10.5mm}
-(1-\frac{p_0}{E_0})v^r({-\bf p},E_0)\otimes\bar v^r({-\bf p},E_0)
\right)
= \, \not\hskip-0.7mm\!{p}\,+\,M,
\label{LR-S} \\
\Lambda_{\alpha\beta}({p},M)
& = &
\frac12 \sum_r \left(
(1+\frac{p_0}{E_0}){\cal U}^r_\alpha({\bf p},E_0)
\otimes\bar {\cal U}^r_{\beta}({\bf p},E_0)
\right. \nonumber \\
& &  \left. \hspace*{10.5mm}
-(1-\frac{p_0}{E_0}){\cal V}^r_\alpha({-\bf p},E_0)
\otimes\bar {\cal V}^r_\beta({-\bf p},E_0)
\right), \\
\Lambda_{\alpha\beta,\gamma\delta}({p},M)
& = &
\frac12 \sum_r \left(
(1+\frac{p_0}{E_0}){\cal U}^r_{\alpha\beta}({\bf p},E_0)
\otimes\bar {\cal U}^r_{\gamma\delta}({\bf p},E_0)
\right. \nonumber \\
& & \left. \hspace*{10.5mm}
-(1-\frac{p_0}{E_0}){\cal V}^r_{\alpha\beta}({-\bf p},E_0)
\otimes\bar {\cal V}^r_{\gamma\delta}({-\bf p},E_0)
\right),
\end{eqnarray}
where $E_0=\sqrt{{\bf p}^2 + M^2}$, and the Rarita-Schwinger spinors read
\begin{eqnarray}
{\cal U}^r_\alpha(p)
& = &
\sum_{\lambda, s} \langle 1\, \lambda \,\frac12\,s|\,\frac32\,
r \rangle \, \varepsilon^\lambda_\alpha(p) \, u^s(p),\\
{\cal U}^r_{\alpha\beta}(p)
& = &
\sum_{\lambda, \lambda' s, t} \langle 1
\, \lambda \,\frac12 \, s| \, \frac32 \, t \rangle \,
\langle \frac32 \, t \, 1\, \lambda' | \, \frac52\, r \rangle\,
\varepsilon^\lambda_\alpha(p) \,
\varepsilon^{\lambda'}_\beta(p) \,
u^s(p).
\end{eqnarray}
The spinors $v$ and ${\cal V}$ are related to $u$ and ${\cal U}$ as
$v(p) = i \gamma_2\, u^*(p)$ and
${\cal V}(p)=i\gamma_2\, {\cal U}^*(p)$, respectively.

The polarization four-vector for a spin-1 particle with spin projection
$\lambda$, four-momentum $p=(E,{\bf p})$ and mass $m_V$ reads
\begin{eqnarray}
\varepsilon^\lambda(p) = \left( \,
\frac{{\bbox \epsilon}^\lambda\cdot{\bf p}}{m_V},\,\,
\frac{{\bf p}\,({\bbox \epsilon}^\lambda\cdot{\bf p})}{m_V ( E + m_V)}\,
\right),
\end{eqnarray}
where the three-dimensional polarization vector $\bbox \epsilon$ is
defined as
\begin{eqnarray}
{\bbox \epsilon}^{\pm 1}=
\mp \frac{1}{\sqrt2} (\,1,\,\,\pm i,\,\,0\,)\qquad
{\bbox \epsilon}^{0}=
(\,0,\,\,0,\,\,1\,).
\end{eqnarray}

In our calculations we use energy-dependent total resonance
decay widths $\Gamma_{N^*}$. However, taking into account that the
effect of a finite width is quite different for $s$ and $u$ channels,
because of the evident relation
$|u|+M^2_{N^*} \gg |s-M^2_{N^*}|$,
we use $\Gamma_{N^*} = \Gamma^0_{N^*}$ for the $u$ channels and
\begin{equation}
\Gamma_{N^*} = \Gamma^0_{N^*}
\left[ 1 - B^\pi_{N^*} + B^\pi_{N^*}\,
\left( \frac{\bf k_{}}{{\bf k}_0} \right)^{2J} \right],
\end{equation}
for the $s$ channels, where $\Gamma^0_{N^*}$ is the total on-shell resonance
decay width and $B^\pi_{N^*}$ stands for the branching ratio
of the $N^*\to N\pi$ decay channel taken from \cite{PDG98};
${\bf k}^{}_0$ is the pion momentum at the resonance position,
i.e.  at $\sqrt{s}=M_{N^*}$,
and the factor
$({\bf k}/{{\bf k}_0})^{2J}$ comes from a direct
calculation of the $N^*\to N\pi$ decay width using the effective Lagrangians
of Eqs.~(\ref{N1440} - \ref{N1720}),
where we keep the leading term proportional to ${\bf k}^{2J}$.
Note, that the effect of the energy dependent finite decay width
is more essential in near-threshold $\omega$ production
for the higher resonances $N_{\frac 32 -}(1700)D_{13}$ and
$N_{\frac 32 -}(1720)P_{13}$, where
$|s-M^2_{N^*}| \sim \Gamma_{N^*}M_{N^*}$, but even there it becomes not so
important because, as we will see, the contribution of these
resonances to the total amplitude
is rather small. Nevertheless, for completeness, we use the above prescription
for all eight considered resonances.

\subsection{Fixing parameters}

The coupling constant $g_{\phi\rho\pi}$ is determined by the
$\phi\to\rho\pi$ decay. The value
$\Gamma_{\phi\to\rho\pi}=0.69$ MeV \cite{PDG98}
results in  $|g_{\phi\rho\pi}|=1.1$ GeV$^{-1}$. The SU(3) symmetry
considerations \cite{Nakayama99,TLTS99} predict a negative value for
it. Thus $g_{\phi\rho\pi}=-1.1$ GeV$^{-1}$.

The coupling constant $g_{\omega\rho\pi}$ is determined by the
$\omega\to\gamma\pi$ decay. Relying on the vector dominance model one gets
$g_{\omega\rho\pi} = 12.9$ GeV$^{-1}$ \cite{SC00}.

The remaining parameters of the meson exchange amplitude for the
process in Fig.~1a are taken from the Bonn model as listed in
Table~B.1 (Model II) of Ref.~\cite{Bonn}:
${g_{\rho NN} = 3.72}$,
${\kappa_{\rho NN} = 6.1}$, and
${\Lambda_{\rho NN} = 1.3}$.
The  parameter $\Lambda^\rho_{V\rho\pi}$ will be determined later.

The nucleon and nucleon resonance amplitudes in Fig.~1b
and Eqs.~(\ref{Ampl1} - \ref{AmplA})
are determined by the couplings
$f_{\pi NN}$,
$f_{\pi NN^*}$,
$g_{\omega NN}$,
$g_{\omega NN^*}$,
$g_{\phi NN}$,
$g_{\phi NN^*}$,
$g_{VNN} \kappa_{VNN}$,
the resonance widths $\Gamma^0_{N^*}$
the branching ratios $B^\pi_{N^*}$,
and the cut-offs $\Lambda_B$.
For the coupling constant $f_{\pi NN}$ we use the standard value
$f_{\pi NN}=1.0$ \cite{RB00,Bonn}.
For the $\omega NN$ coupling we use the value
$g_{\omega NN} = 10.35$ determined recently \cite{RSY99}
by fitting the nucleon-nucleon scattering data. This value as well as
$\kappa_{\omega NN} = 0$ is close to the one which has been found
in a study of $\pi N$ scattering and the reaction
$\gamma N \to \pi N$ \cite{SL96}.

The values of coupling constants $f_{\pi NN^*}$ are determined from
a comparison of calculated decay widths $\Gamma_{N^*\to N\pi}$
with the corresponding experimental values \cite{PDG98}.
The corresponding signs are taken in accordance
with the quark model prediction of Ref.~\cite{RB00}.

The values of coupling constants $g_{\omega NN^*}$ are found as
$g_{\omega NN^*} = [g_{\omega NN^*}/g_{\omega NN}] g_{\omega NN}$,
where the ratio $[g_{\omega NN^*}/g_{\omega NN}]$ is determined by
the quark model calculation of Ref.~\cite{RB00}. For convenience we
show all the coupling constants, decay widths and branching ratios
used in our calculation in Table~1.
The masses, decay widths and branching ratios in Table~1
represent the averages in \cite{PDG98}.
Slightly different values have
been extracted in \cite{Mosel_Haberzettel} in a recent reanalysis
of the data.

The couplings $\phi NN$,  $\phi NN^*$
determined by SU(3) symmetry considerations are
\begin{eqnarray}
g_{\phi NN} =-{\rm tg} \Delta\theta_V\, g_{\omega NN},
\qquad g_{\phi NN^*} =-{\rm tg} \Delta\theta_V\, g_{\omega NN^*},
\end{eqnarray}
where $ \Delta\theta_V \simeq 3.7^0$
is the deviation from the ideal $\omega - \phi$ mixing.
Similarly, we assume
$g_{\phi NN} \kappa_{\phi NN} \simeq -{\rm tg} \Delta\theta_V \,
g_{\omega NN} \kappa_{\omega NN} = 0$, or
$\kappa_{\phi NN} \simeq 0$,
which is consistent with the estimate in \cite{MMSV}.

The yet undetermined parameters are: the cut-off parameters
$\Lambda^\rho_{\phi\rho\pi}$ and  $\Lambda^\rho_{\omega\rho\pi}$
for the virtual $\rho$ meson in the $V\rho\pi$ vertex,
the cut-off $\Lambda_N$ and the eight cut-offs $\Lambda_{N^*}$
in Eq.~(\ref{cutN}).
We can reduce the number of parameters by making the natural assumptions
\begin{eqnarray}
\Lambda^\rho_{\phi\rho\pi}&=&
\Lambda^\rho_{\omega\rho\pi}\equiv\Lambda^\rho_V ,
\label{LV}\\
\Lambda_N &=& \Lambda_{N^*}\equiv\Lambda_B.
\label{LB}
\end{eqnarray}
Our best fit of total cross sections of existing data
is obtained by $\Lambda^\rho_V=1.24$ GeV and $\Lambda_B=0.66$ GeV.

\section{Results}

The results of our full calculation of the total cross sections
as a function of the energy excess $\Delta s^{1/2}=\sqrt{s}-M_N-m_V$,
including all amplitudes depicted in
Fig.~1, are represented by the solid curves in Fig.~2. We also show separately
the contributions of meson exchange, nucleon and nucleon resonance channels.
The data for the reaction
$\pi^-p\to\phi n$ are taken from Ref.~\cite{DATAome_p},
while the data for the reactions
$\pi^+n\to\omega p$ and  $\pi^-p\to\omega n$
are from Refs.~\cite{DATAome_p,DATAome_m}, respectively.
Note that here we display the total cross section of the reaction
$\pi^-p\to\omega n$, $\sigma_{\rm tot}$,
which differs from the differential
cross section  $\sigma_{\rm dif}$ in Ref.~\cite{DATAome_m}
by a factor~\cite{SC00,HK00} included in the phase space of the
unstable $\omega$ meson,
\begin{eqnarray}
\sigma_{\rm tot} = \sigma_{\rm dif}
\left\{
\int\limits_{\sqrt{ {P}^2_{\rm min} +M_N^2}}\limits^{\sqrt {{P}^2_{\rm max} +M_N^2}}
\sqrt{\frac{\lambda_f(s')\lambda_i(s)}{\lambda_f(s)\lambda_i(s')}}\,
\frac{2\sqrt{s'}\Gamma_\omega m_\omega\,dE}{\pi((m_\omega^2-s'+2\sqrt{s'}E-M_N^2)^2
+ \Gamma_\omega^2 m_\omega^2)} \right\}^{-1} ,
\end{eqnarray}
where $\sqrt{s'}=E+\sqrt{E^2-M_N^2+m^2_\omega}$, $\Gamma_\omega$ is the
$\omega$ decay width and $\lambda_i(s)\,(\lambda_f(s))
=\lambda(s,m_\pi^2,M_N^2)\,(\lambda(s,m_\omega^2,M_N^2))$ with
$\lambda(x,y,z)=(x-y-z)^2-4yz$. The intervals $[P_{\rm max},P_{\rm min}]$
for given $p'$ (or $s$) are as in \cite{DATAome_m}.

From Fig.~2 (left panel)
it is evident that the total amplitude of $\omega$ production
is a result of
strong interferences of all channels: the meson exchange (dot-dashed curve),
the nucleon term (long dashed curve) and the resonance contribution (dashed
curve) play a comparative role.
For the $\phi$ production (cf.\ Fig.~2, right panel)
only meson and nucleon terms are
important. The resonance contribution is rather small and is therefore
not displayed here. Moreover the relative contribution of the nucleon term in
$\phi$ production is much smaller than for $\omega$ production. That
is because the initial energy $\sqrt{s}= M_N + m_V + \Delta s^{1/2}$
is greater for the $\phi$ production at the same energy  excess
and as a consequence, we have two suppression factors:
(i) the nucleon/resonance denominators and
(ii) the form factors $F_{N, N^*}$ in Eqs.~(\ref{Ampl1} - \ref{AmplA}).
The interference of the meson exchange and nucleon terms is almost
destructive, while the contribution of the resonant part is more
complicated because the amplitude is complex with different phases
for different resonances.

In order to illustrate the structure of the resonant part
we show in Fig.~3 the contribution of each resonance separately
as a function of the $\omega$ production angle at two excess energies
$\Delta s^{1/2}=20$ (100) MeV at left (right) panel.
One can see that just near the threshold the resonances with
$J = \frac12$ are important. Together with the nucleon term they are
$N_{\frac 12^+}(1440) P_{11}$,
$N_{\frac 12^-}(1535) S_{11}$ and
$N_{\frac 12^-}(1650) S_{11}$. It is interesting that the separate
contributions of the two latter ones are greater than the nucleon term.
But their phases are opposite and, therefore, they cancel each other.
The cancellation increases with energy which results in a total
decrease of the resonance contribution.
On the other hand one can see that the relative role of the
higher spin resonances with orbital/radial excitations, being proportional
to ${\bf q}^2$ and ${\bf q}^4$, increases with increasing values of
$\Delta s^{1/2}$, as illustrated in Fig.~4.
This enhancement, however, is smaller than the effect of
the strong destructive interference
of the  resonance amplitudes  and, therefore, the total contribution
of the resonance channel decreases with energy as shown in Fig.~2.

In our analyses we do not include the *** resonance
$N_{\frac 12 +}(1710)P_{11}$ \cite{PDG98}, as in \cite{RB00}.
A simple estimate shows that its contribution is rather small. Indeed,
the calculation in \cite{PM00}, within the vector dominance model,
shows that the
$\omega NN^*$ coupling for
$N_{\frac 12 +}(1710)P_{11}$ is about four times smaller than
for the
$N_{\frac 12 +}(1440)P_{11}$ resonance.
Observing further that the corresponding pion decay width of
$N_{\frac 12 +}(1710)P_{11}$,
5$\cdots$15 MeV, is much smaller than that of
$N_{\frac 12 +}(1440)P_{11}$,
210$\cdots$250 MeV, we can expect a negligible
contribution of $N_{\frac 12 +}(1710)P_{11}$
being about one order of magnitude smaller than
the one of $N_{\frac 12 +}(1440)P_{11}$.

Figs.~5 and 6 show the angular distributions of the
$\omega$ and $\phi$ production cross sections
at $\Delta s^{1/2} = 20$ and $100$ MeV,
respectively.
One can see that the destructive interference
between meson exchange and nucleon
channels, which is stronger for backward production, results in
a non-monotonic angular distribution. The effect is stronger at 
$\Delta s^{1/2} = 100$ MeV.

Fig.~7 (left panel) shows the ratio of the averaged amplitudes
$|T_V|$ of
$\omega$ and $\phi$ production as a function of the vector meson's production
angle at $\Delta s^{1/2}=20$ MeV. $|T_V|$ is defined by
\begin{equation}
|T_V| = \left[
{\sum_{m_i,m_f,\lambda}|T^V_{m_f,\lambda;m_i}|^2} \right]^{\frac12},
\end{equation}
where $m_i,m_f,\lambda$ are the spin projection of target, recoil
protons and vector meson, respectively.
The dotted straight lines in Fig.~7
correspond to the standard OZI rule violation value
$R^{OZI}_{\omega/\phi} =
{\rm ctg}\Delta \theta_V = 15.43$.
The long dashed curve corresponds to the ratio of pure nucleon
channels taken separately,
the dot-dashed curve is result for a pure meson exchange,
while the solid line represents the full calculation.
Note that the ratio even for pure meson exchange
amplitudes $R^M_{\omega/\phi}$ is smaller than $R^{OZI}_{\omega/\phi}$.
That is because at the threshold
\begin{eqnarray}
R^M_{\omega/\phi}\simeq
\frac{g_{\omega\rho\pi}}{g_{\phi\rho\pi}}\,
\frac{m_\omega}{m_\phi}\,
\frac{f(m_\omega)}{f(m_\phi)}\simeq 9.9\,
\frac{f(m_\omega)}{f(m_\phi)},
\end{eqnarray}
where $f(m)$ is a smooth function of $m$.
The ratio for pure
resonance terms is greater than $R^{OZI}_{\omega/\phi}$ by
an order of magnitude and more, i.e.,
$R^{N^*}_{\omega/\phi}\sim 500\,\,(250)$
at $\theta=\pi\,\,(0)$,
because of a strong propagator and form factor suppression for $\phi$
production. In the absence of the resonant amplitude the
destructive interference
of meson exchange (M) and nucleon (N)
channels results in  $R^{M+N}_{\omega/\phi}<
R^M_{\omega/\phi}$.  The presence of the resonance components leads
to $R^M_{\omega/\phi} < R_{\omega/\phi} < R^{OZI}_{\omega/\phi}$.

Fig.~7 (right panel) shows the ratio of the
angular integrated amplitudes
\begin{equation}
<|T_V|>_\Omega=\left[
\sum_{m_i,m_f,\lambda}
\frac{1}{4\pi}\int d\Omega \, |T^V_{m_f,\lambda;m_i}|^2
\right]^{\frac12}
\end{equation}
of $\omega$ and $\phi$ production as a function of $\Delta s^{1/2}$.
One can see that  $R_{\omega/\phi}$ (solid curve)
may be slightly above or below
the pure $R^{M}_{\omega/\phi}$ value (dot-dashed curve),
however, remaining much smaller than $R^{OZI}_{\omega/\phi}$,
namely ${R_{\omega/\phi} = 7.5 \cdots 10}$,
in agreement with the analysis in \cite{SC00}.

Fig.~8 shows the results of our full calculation of the spin density matrix
element $\rho_{00}$ at $\Delta s^{1/2}=20$ and 100 MeV in the left and
right panels, respectively. Near the threshold the meson exchange amplitude
behaves as
\begin{eqnarray}
T^{(M)}_\lambda \sim {\bf k}\cdot [{\bf k}
\times {{\bbox \epsilon}^*}^\lambda].
\label{TM}
\end{eqnarray}
That means that only $\lambda = \pm 1$ contributes and,
therefore, $\rho_{00}$ is suppressed.
The pure nucleon $s$ channel amplitude behaves as
\begin{eqnarray}
T^{(N)}_\lambda\sim
\langle f|\,\bbox \sigma \cdot {{\bbox \epsilon}^*}^\lambda |i\rangle,
\label{TN}
\end{eqnarray}
which results in an isotropic spin density, i.e.,
$\rho_{00}=\rho_{11}=\rho_{-1-1} = 1/3$.
The resonance amplitudes have additional terms proportional
to $ {\bf k} \cdot {{\bbox \epsilon}^*}^\lambda$, which also enhance
$\rho_{00}$. This effect is seen clearly in the left panel
of Fig.~8
for which our qualitative analysis is valid. For $\phi$ production,
where the main contribution comes from the meson exchange channel
(cf.\ Figs.~2 and 5),
$\rho_{00}$  is relatively small, $\rho_{00}< 0.05$. But for $\omega$
production, where the contribution of the resonance channel is essential,
we find  $\rho_{00}\sim 0.3$ which is close to an isotropic spin-density
distribution with $\rho_{00} \simeq \rho_{11} = \rho_{-1-1} \simeq 1/3$.

Fig.~9
shows the angular distribution of hadronic decays
$\phi\to K^+K^-$ and $\omega\to \pi^+\pi^-\pi^0$ in
$\pi^-p\to n V$ reactions at  $\Delta s^{1/2}=20$ MeV.
The left panel corresponds to a calculation
of the vector meson production in forward direction, $0.9 < \cos\theta <1$,
where the cross section achieves a maximum, while the right
panel shows its average value in the full angular interval
$-1 < \cos \theta < 1$. One can see a
striking difference between the $\phi$ and $\omega$ cases:
an almost anisotropic distribution for $\phi$ production
(${\cal W} \simeq \frac32 \, \sin^2 \Theta$)
and almost isotropic distribution for $\omega$
production (${\cal W}\sim1$), respectively,
which reflect the difference in the corresponding
production mechanisms.

A similar difference is predicted for the
angular distribution of electrons in
$\pi^-p \to n V \to n e^+e^-$ reactions shown in Figs.~10 and 11
at  $\Delta s^{1/2}=20$ and 100 MeV, respectively.
Again one can see that at $\Delta s^{1/2} = 20$ MeV
(Fig.~10)
the $\phi$ and $\omega$ cases are very different:
an almost anisotropic distribution for $\phi$ production
(${\cal W} \simeq \frac34 (1 + \cos^2 \Theta)$),
and almost isotropic distribution for $\omega$ production
(${\cal W} \sim 1$), respectively.
This pronounced difference disappears at $\Delta s^{1/2} = 100$ MeV
(Fig.~11).

It should be emphasized
that our prediction for the separate decay $\omega \to e^+e^-$
may be tested experimentally supposed the corresponding
detector acceptance is sufficiently large
for distinguishing the sharp $\omega$ resonance peak
in the dielectron invariant mass distribution
sitting on the background of the wide
$\rho^0$ meson contribution. Otherwise one should consider
the $\omega-\rho^0$ interference as calculated
for unpolarized observables in \cite{SLF00}.
However,
having in mind that the $\rho^0$ production is a competing channel to
the $\omega$ production, where the role of the resonances is expected
to be even more important because of the larger number of intermediate
$N^*$ states, we expect that our prediction of an almost
isotropic $e^+e^-$  distribution at invariant mass
$M_{e^+e^-} \simeq m_\omega$
remains valid in general,
reflecting the role of the
baryon resonances in the production mechanism.

\section{Summary}

In summary we have performed a combined analysis of $\omega$ and $\phi$
production in $\pi N$ reactions near the threshold at the same energy
excess. We find that the meson-exchange amplitude alone can not
describe the existing data, rather the role of the direct nucleon term and the
nucleon resonance amplitudes is essential.
The latter statement is very important for the
$\omega$ production, and therefore we investigate the role
of nucleon resonances in detail. The $\omega NN^*$ couplings as well as
the phases of the $\pi NN^*$ couplings are taken from the recent work
\cite{RB00}. It is found that the resonance contributions
can influence significantly the total and the differential cross
sections at small energy excess as well as the ratio of the averaged
amplitudes of $\omega$ and $\phi$ production.  For this ratio
we get the value
$8.7 \pm 1.5$
which is much smaller than the value based
on the standard OZI rule violation.
The dominant contributions are found to stem from
the nucleon resonances
$N_{\frac 12 -}(1535)S_{11}$,
$N_{\frac 12 -}(1650)S_{11}$, and
$N_{\frac 12 +}(1440)P_{11}$.
However, the other resonances become also important
with increasing energy excess.

We have shown that our predictions can essentially be
tested by measuring the angular distribution of decay particles
in the reactions $\pi N\to N \phi \to N K^+K^-$,
$\pi N\to N \omega \to N 3\pi$ and
$\pi N\to N V \to N e^+e^-$. 
Near the thresholds,
for the $\phi$ production we
predict an almost anisotropic distribution,
while for the $\omega$ production an almost
isotropic distributions is obtained.
Experimentally, this prediction can be tested
with the pion beam at the HADES spectrometer at GSI/Darmstadt \cite{HADES}.

It should be stressed that the present investigation is completely
based on the conventional meson-nucleon dynamics and,
therefore, our predictions may be considered as a necessary
background for forthcoming studies of the hidden strangeness
degrees of freedom in non-strange hadrons.
Finally, it should be emphasized that our study here is a
first step from the point of view of a dynamical treatment of the
problem, thus going beyond \cite{SC00}.
The main uncertain part is the poor knowledge of the
strong cut-off factors for the virtual nucleon
and nucleon resonances
which are important for the present consideration.
Also, the effects of the final state interaction must
be investigated which may be pursued by extending the approach of
Refs.~\cite{TKR00,SS68}.


\subsection*{Acknowledgments}

We gratefully acknowledge  fruitful discussions with H.W. Barz,
R. Dressler, S.B. Gerasimov, L.P. Kaptari, and J. Ritman.
One of the authors (A.I.T.) thanks for the warm hospitality
of the nuclear
theory group in the Research Center Rossendorf.
This work is supported by BMBF grant 06DR921,
Heisenberg-Landau program,
HADES-JINR participation project \#03-1-1020-95/2002,
and Russian Foundation for Basic Research under Grant No. 96-15-96426.


\begin{table}
\centering
\begin{tabular}{cccccc}

baryon & $M_{N^*} $ & $f_{\pi NN^*}$ & $g_{\omega NN^*}$ &
$\Gamma^0_{N^*} $ &$B^{\pi}_{N^*} $  \\ \hline
$N\frac12^+\,N$ & $940$ & $1.0$   & $10.35$  & --- & ---\\ \hline
$N\frac12^+\,P_{11}$ & $1440$ & $0.39$    & $6.34$  &$350$   &$0.65$\\ \hline
$N\frac32^-\,D_{13}$ & $1520$ & $-1.56$   & $8.88$  &$120$   &$0.55$\\ \hline
$N\frac12^-\,S_{11}$ & $1535$ & $0.36$    & $-5.12$  &$150$  &$0.45$\\ \hline
$N\frac12^-\,S_{11}$ & $1650$ & $0.31$    & $2.56$  &$150$   &$0.73$\\ \hline
$N\frac52^-\,D_{15}$ & $1675$ & $0.10$    & $10.87$  &$150$   &$0.45$\\ \hline
$N\frac52^+\,F_{15}$ & $1680$ & $-0.42$    & $-14.07$  &$130$  &$0.65$\\ \hline
$N\frac32^-\,D_{13}$ & $1700$ & $0.36$    & $2.81$  &$100$   &$0.10$\\ \hline
$N\frac32^-\,P_{13}$ & $1720$ & $-0.25$    & $-3.17$  &$150$  &$0.15$ \\
\end{tabular}
\vspace{0.5cm}
\caption{
Parameters for the resonance masses, coupling constants,
total decay widths and branching ratios for $N^*\to N\pi$ decays.
The resonance masses and decay widths are in units of MeV.}
\label{tab:Nstar1}
\end{table}


\vspace*{9mm}

\begin{figure}
\centering
{\epsfig{file=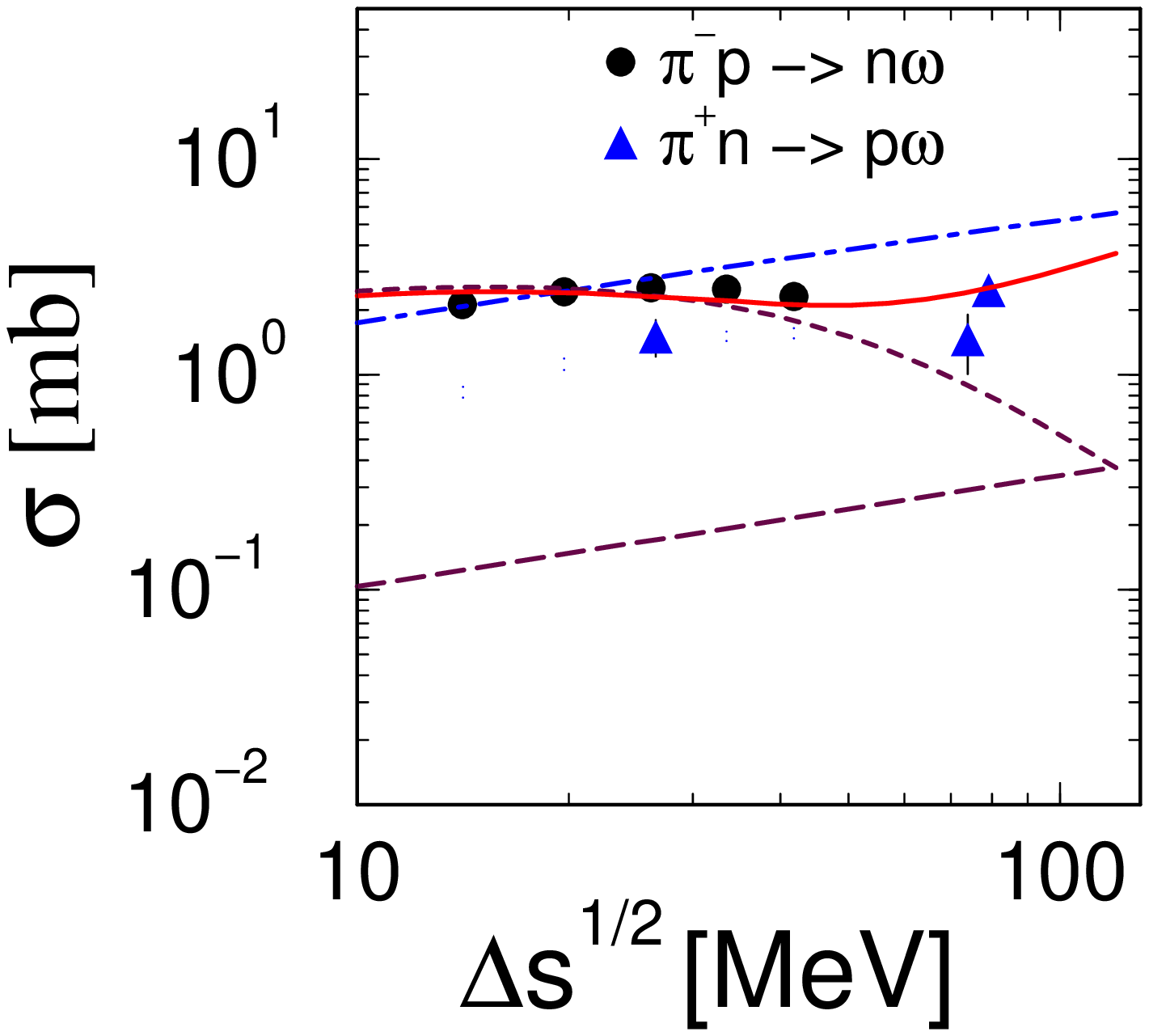, width=7.3cm}\qquad
 \epsfig{file=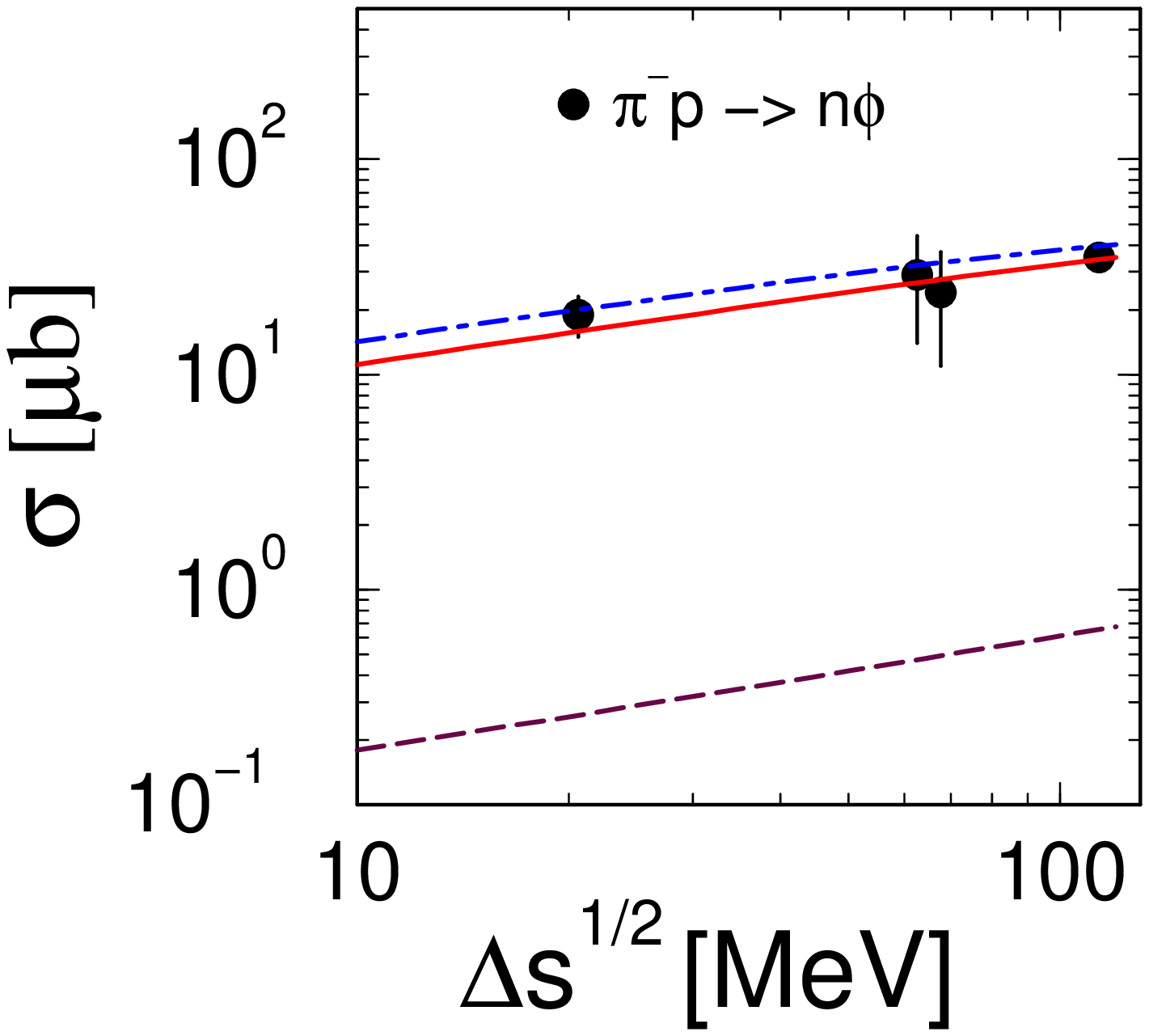, width=7.3cm}}
\caption{
Total cross sections for the reactions
$\pi N\to N\omega$ (left panel) and
$\pi^- p\to n\phi$ (right panel)
as a function of the energy excess $\Delta s^{1/2}$.
The meaning of the curves is:
meson exchange - dot-dashed,
direct and crossed nucleon terms - long-dashed,
$N^*$ resonances -  dashed,
full amplitude - solid.
Data from \protect\cite{DATAome_p,DATAome_m}.}
\label{fig:2}
\end{figure}

\begin{figure}
\centering
{\epsfig{file=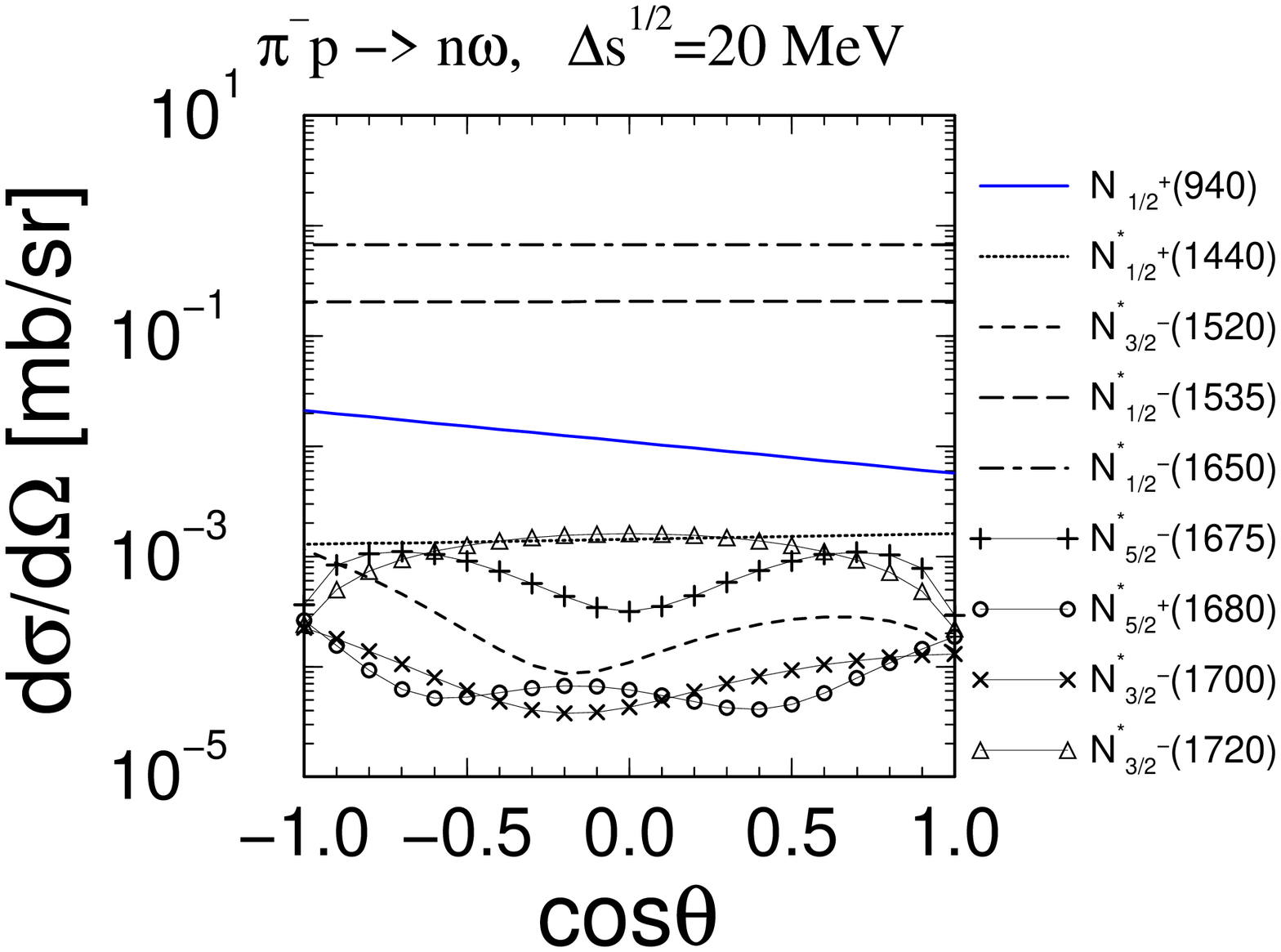, width=7cm}\qquad\qquad
 \epsfig{file=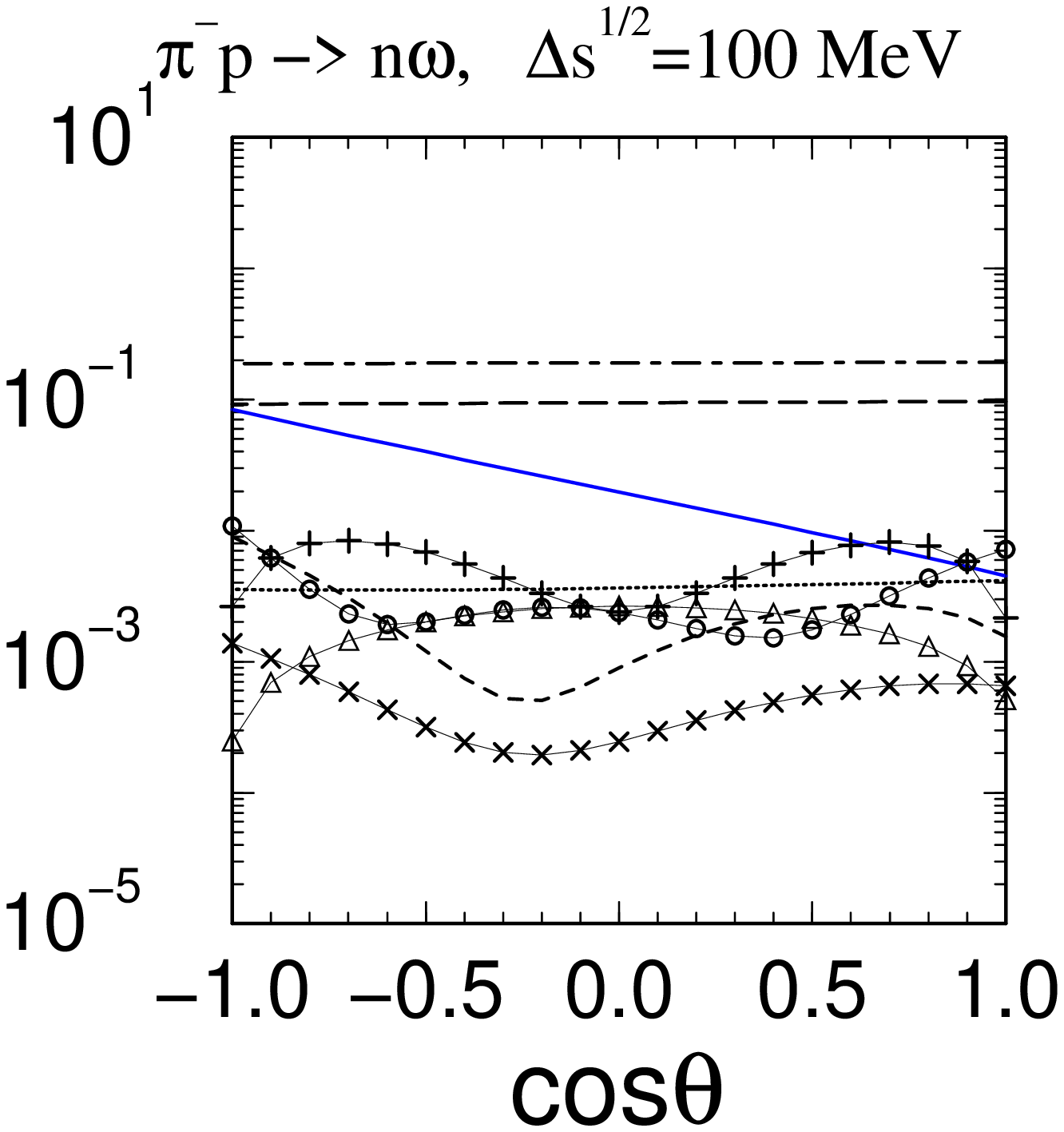, width=6.5cm}}
\caption{
Individual contributions of nucleon resonances listed in Table~1
to the angular differential cross section of $\omega$ production
at  $\Delta s^{1/2} = 20$ MeV (left panel)
and 100 MeV (right panel).}
\label{fig:3}
\end{figure}

\begin{figure}
\centering
{\epsfig{file=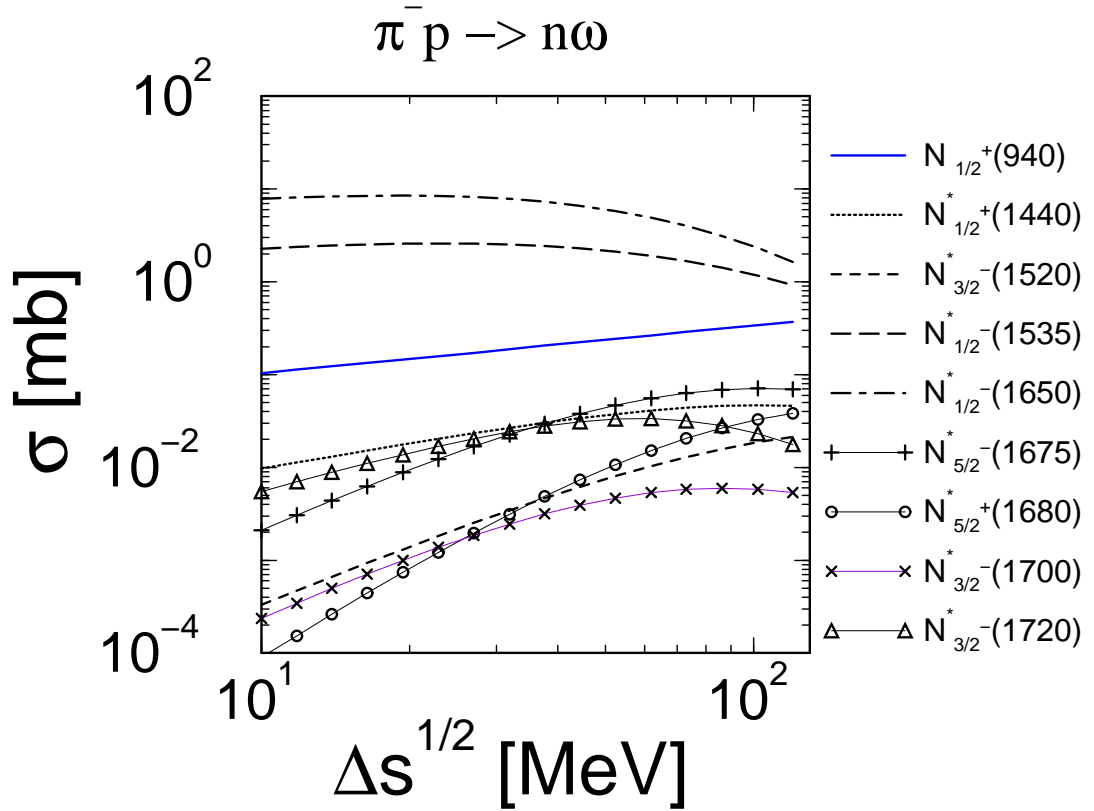, width=12cm}}
\caption{
Individual contributions of nucleon resonances listed in Table~1
to the total cross section of $\omega$ production as a function
of $\Delta s^{1/2}$.}
\label{fig:4}
\end{figure}

\newpage
\begin{figure}
\centering
{\epsfig{file=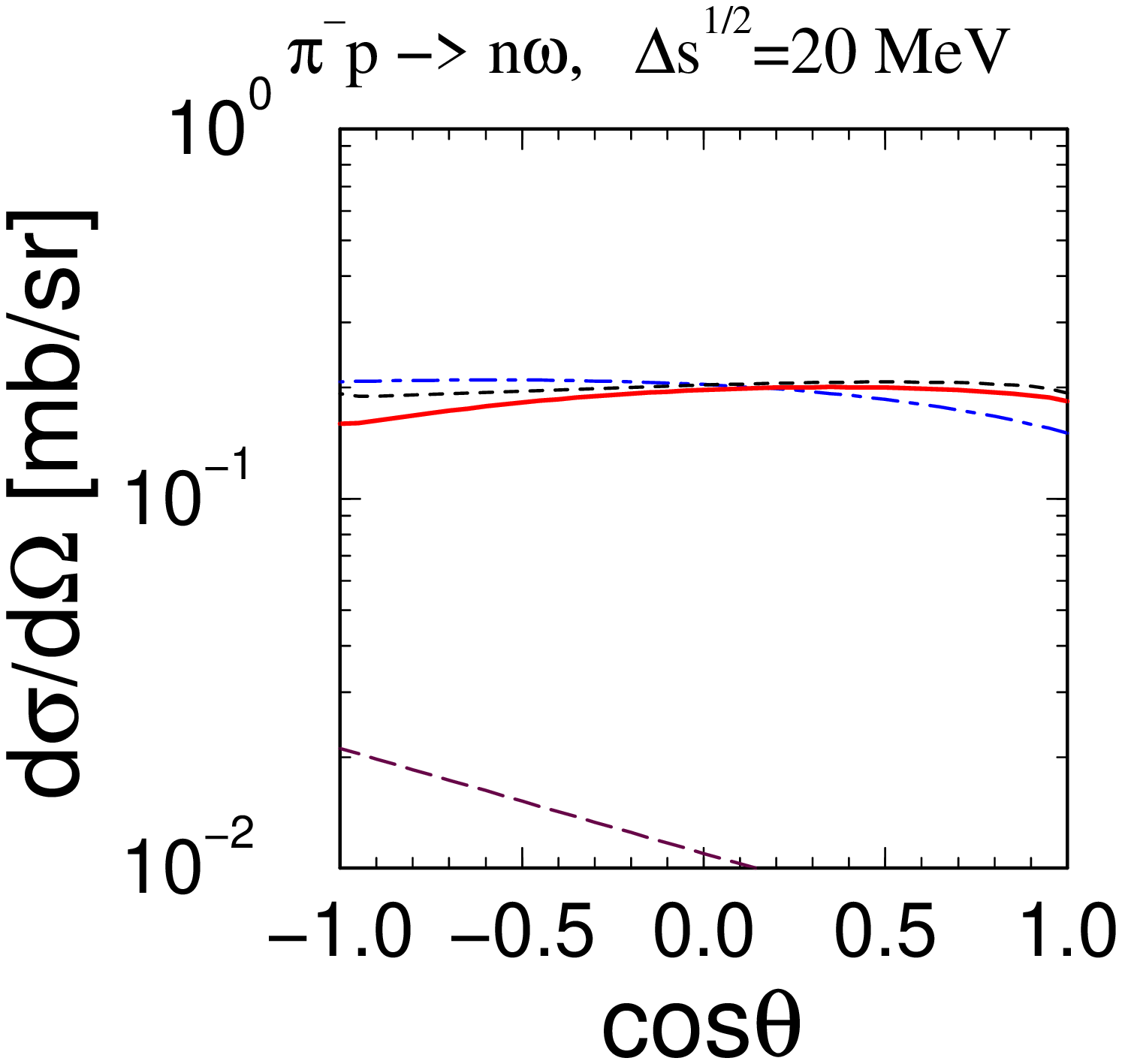, width=7.00cm}\qquad
 \epsfig{file=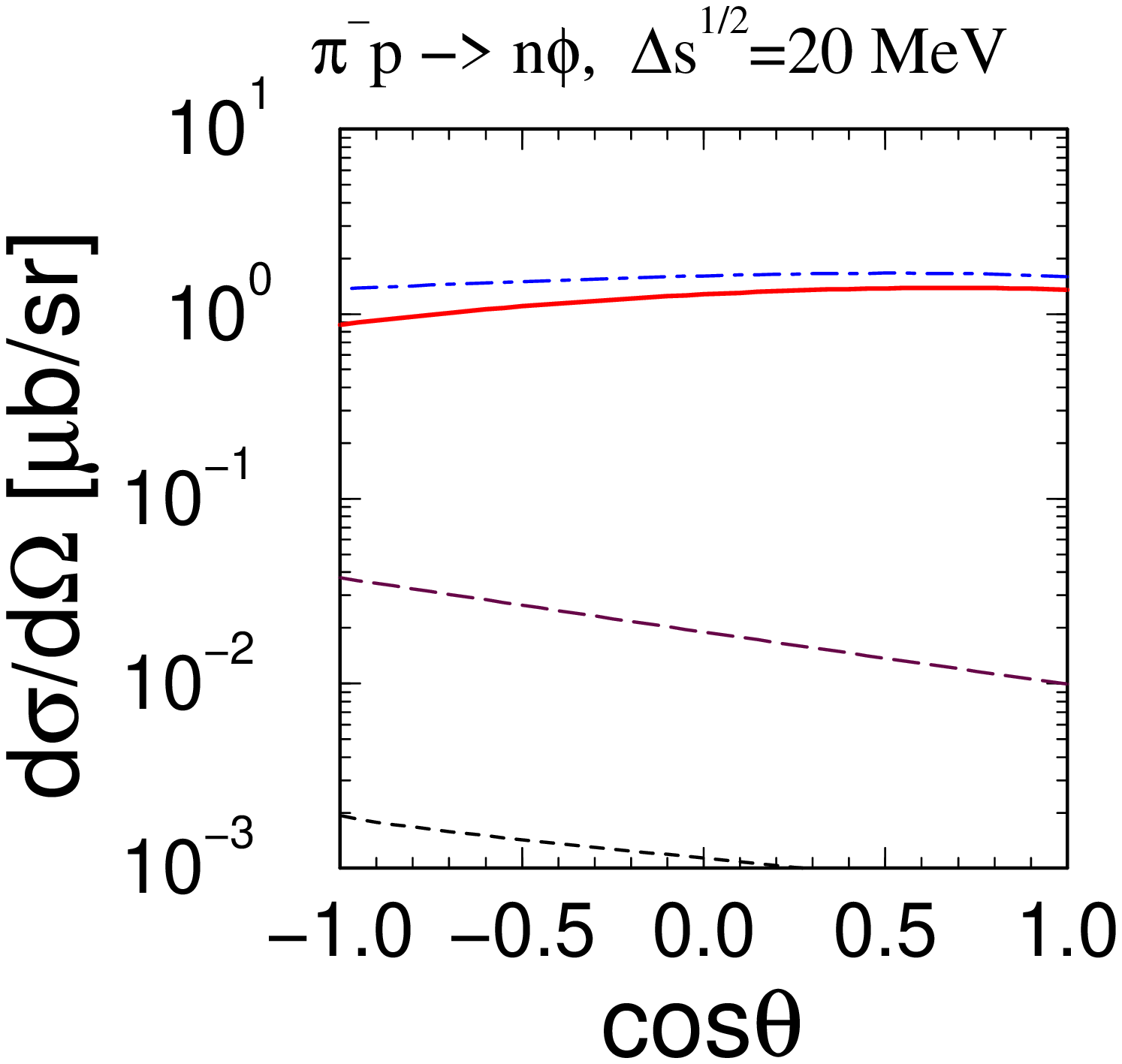, width=7.00cm}}
\caption{
Angular differential cross sections for the reactions
$\pi^- p\to n \omega$ (left panel) and
$\pi^- p\to n\phi$ (right panel)
at $\Delta s^{1/2}=20$ MeV.
Notation as in Fig.~2.}
\label{fig:5}
\end{figure}

\vspace*{9mm}

\begin{figure}
\centering
{\epsfig{file=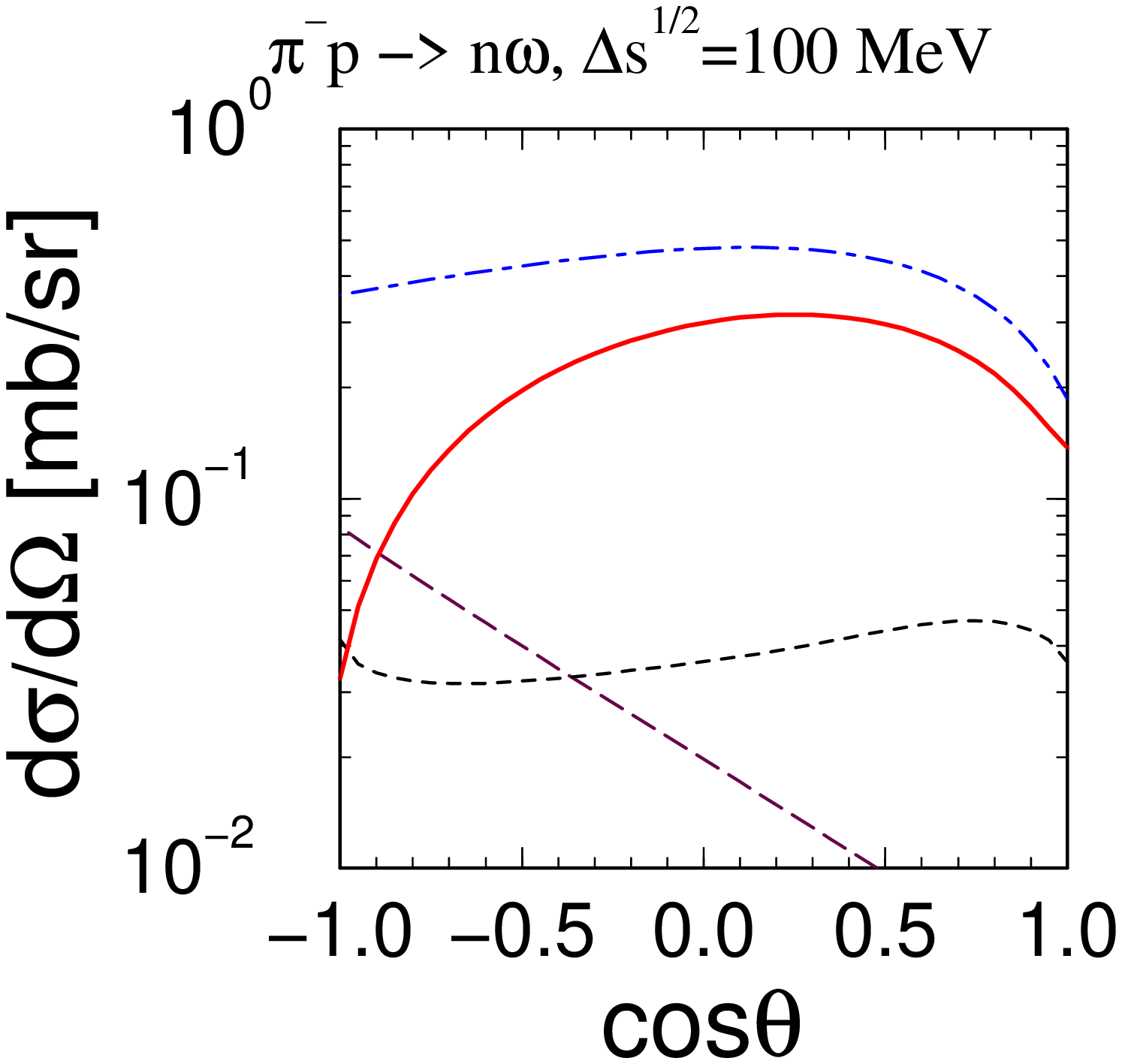, width=7.1cm}\qquad
 \epsfig{file=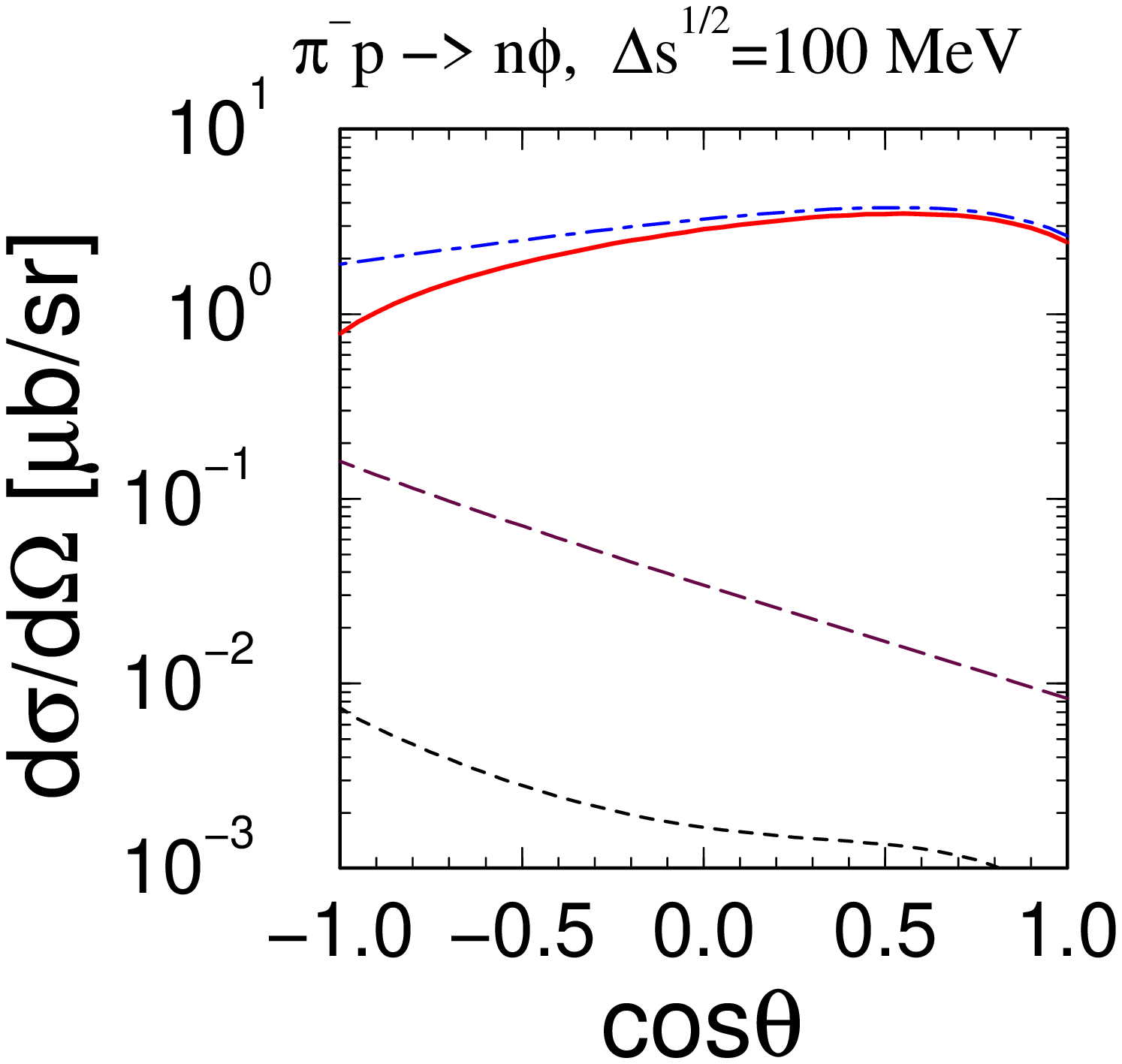, width=7.0cm}}
\caption{
As in Fig.~5 but at $\Delta s^{1/2} = 100$ MeV.}
\label{fig:6}
\end{figure}

\newpage
\begin{figure}
\centering
{\epsfig{file=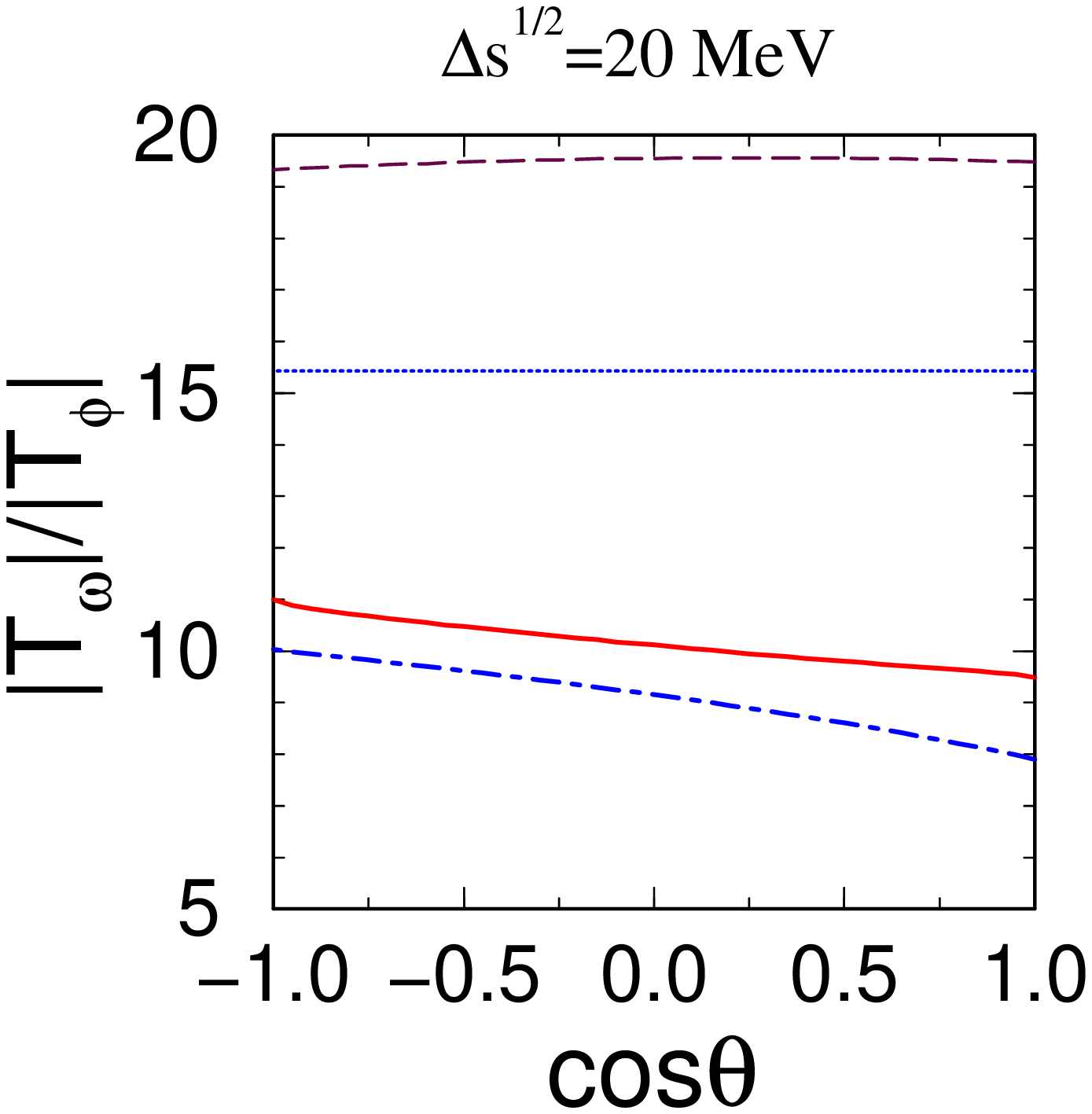, width=7cm}\qquad\qquad
 \epsfig{file=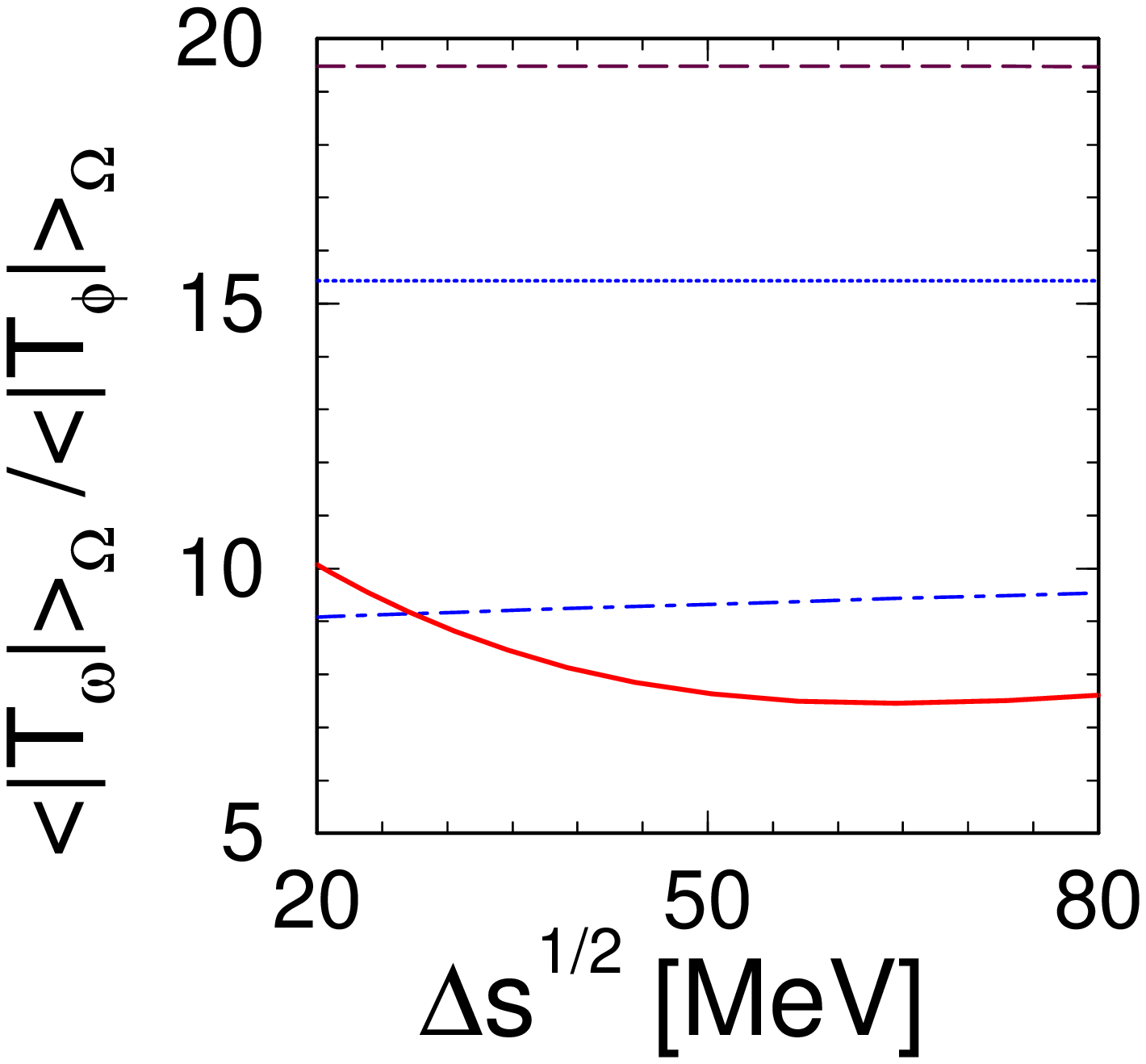, width=7.2cm}}
\caption{
Ratio of the amplitudes of $\omega$ and $\phi$ production.
Left panel: the ratio as a function of 
$\cos \theta$ at $\Delta s^{1/2} = 20$ MeV,
right panel: the ratio averaged over production angle
as a function of $\Delta s^{1/2}$.
Notation as in Fig.~2.}
\label{fig:7}
\end{figure}

\vspace*{9mm}

\begin{figure}
\centering
{\epsfig{file=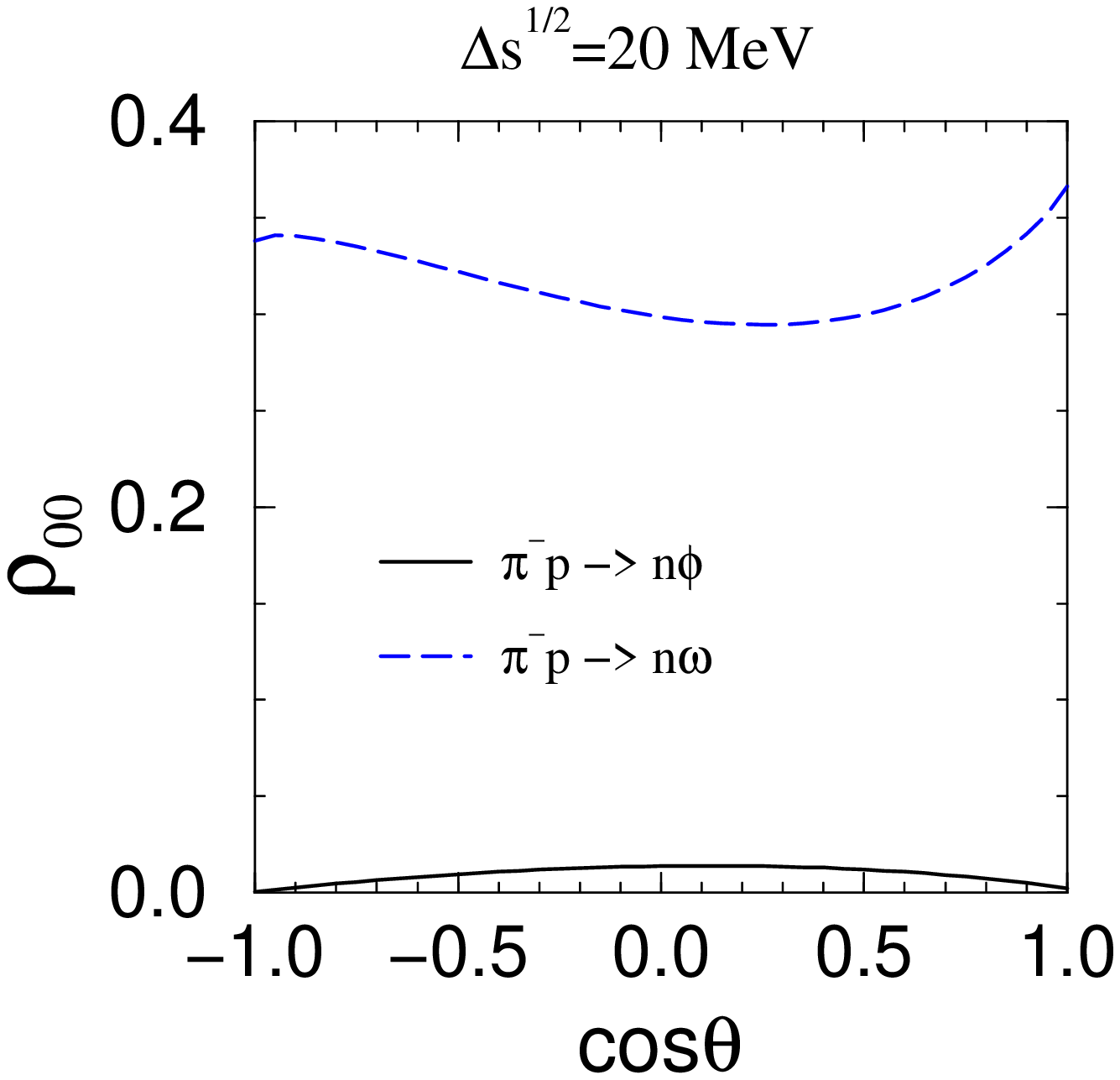, width=7cm}\qquad
 \epsfig{file=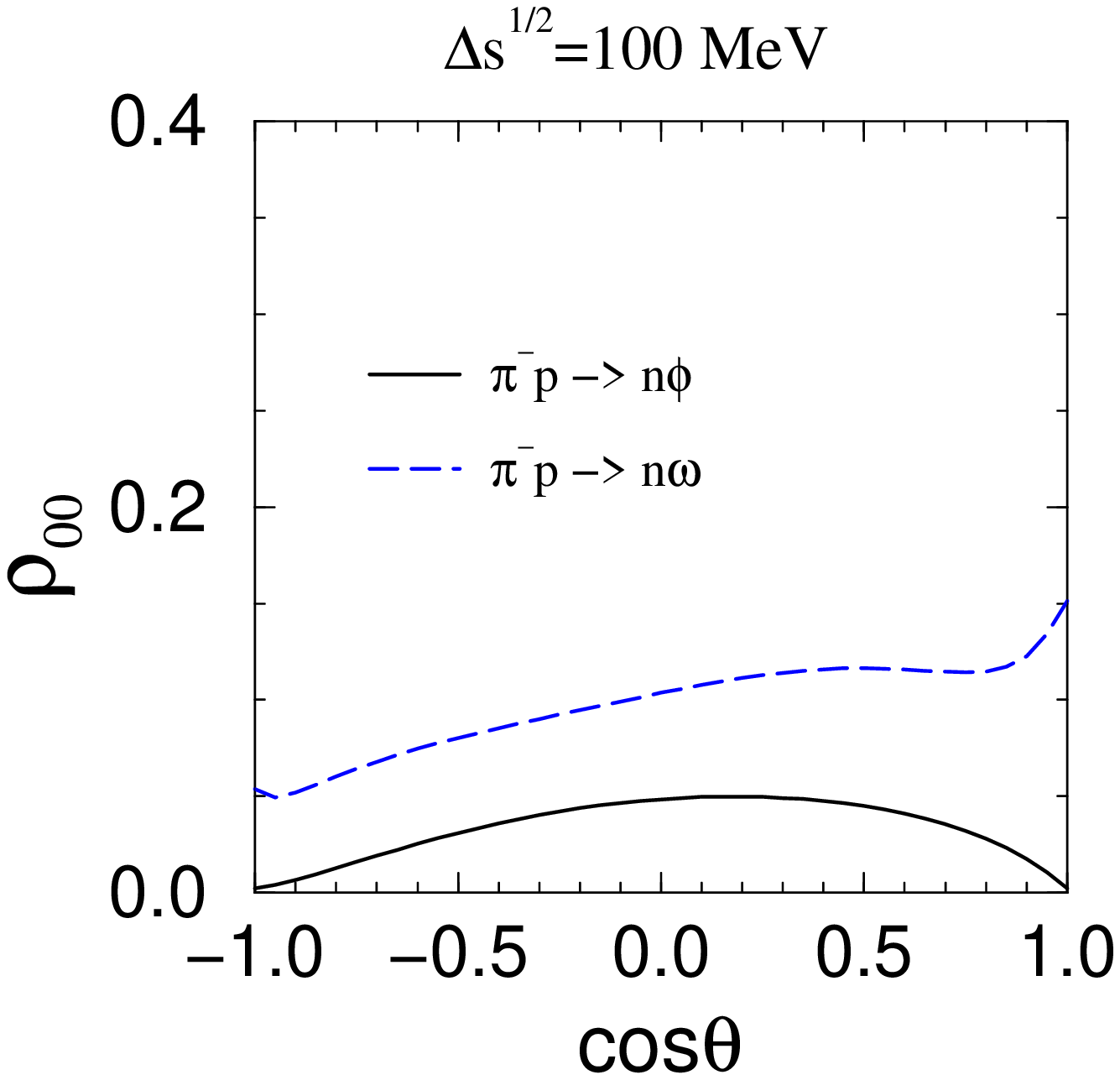, width=7cm}}
\caption{
Spin-density matrix element $\rho_{00}$ for $\omega$ and $\phi$
production as a function
of $\cos \theta$ at $\Delta s^{1/2} = 20$ MeV (left panel)
and 100 MeV (right panel).}
\label{fig:8}
\end{figure}

\newpage
\begin{figure}
\centering
{\epsfig{file=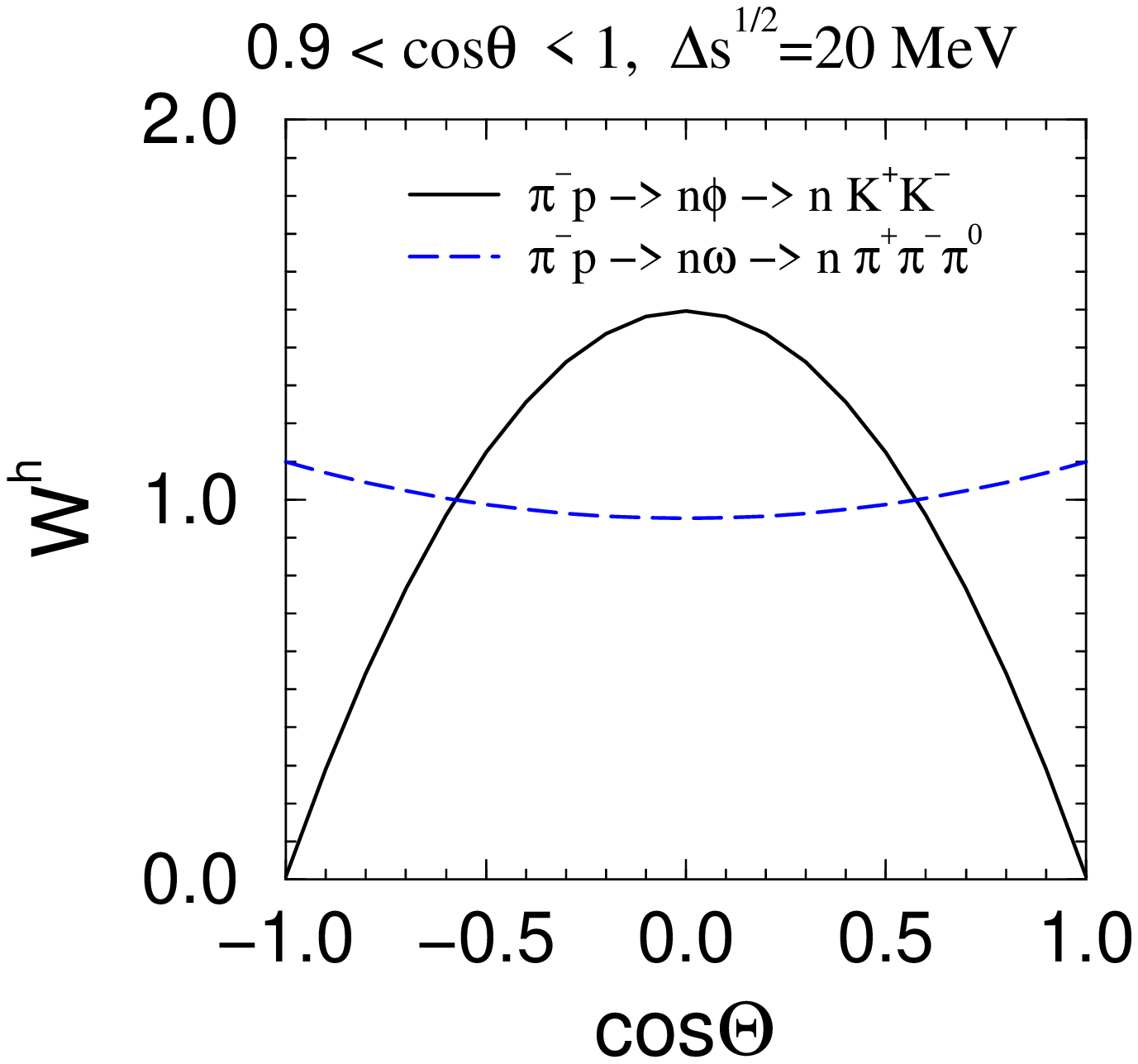, width=7cm}\qquad
 \epsfig{file=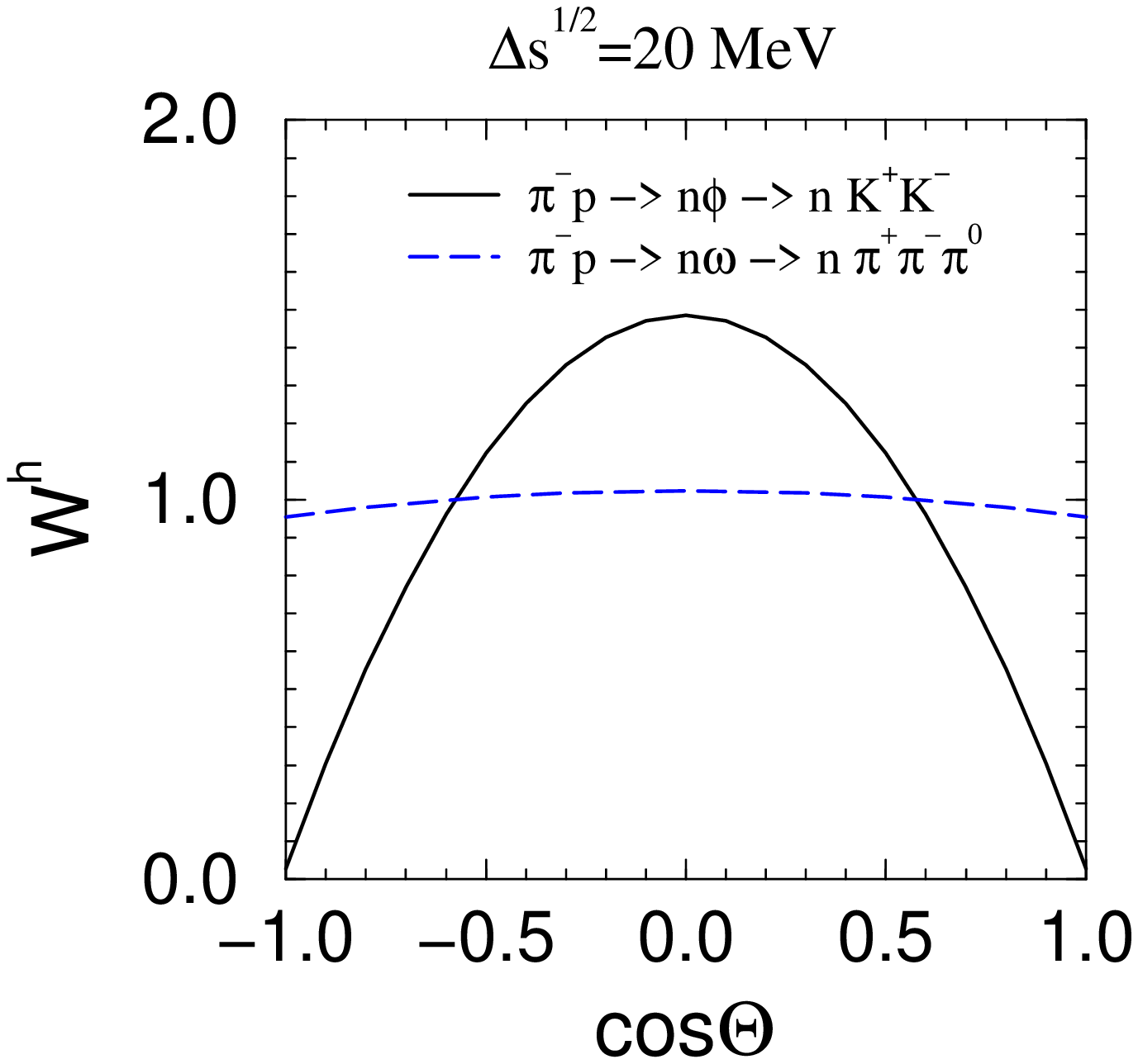, width=7cm}}
\caption{
Meson angular distributions in the reactions
$\pi^- p \to n \phi \to n K^+K^-$ and
$\pi^- p \to n \omega \to n \pi^+\pi^-\pi^0$,
at $\Delta s^{1/2} = 20$ MeV.
Left panel: the distribution at forward vector meson production angles,
right panel: the distribution averaged over all production angles.}
\label{fig:9}
\end{figure}

\vspace*{9mm}

\begin{figure}
\centering
{\epsfig{file=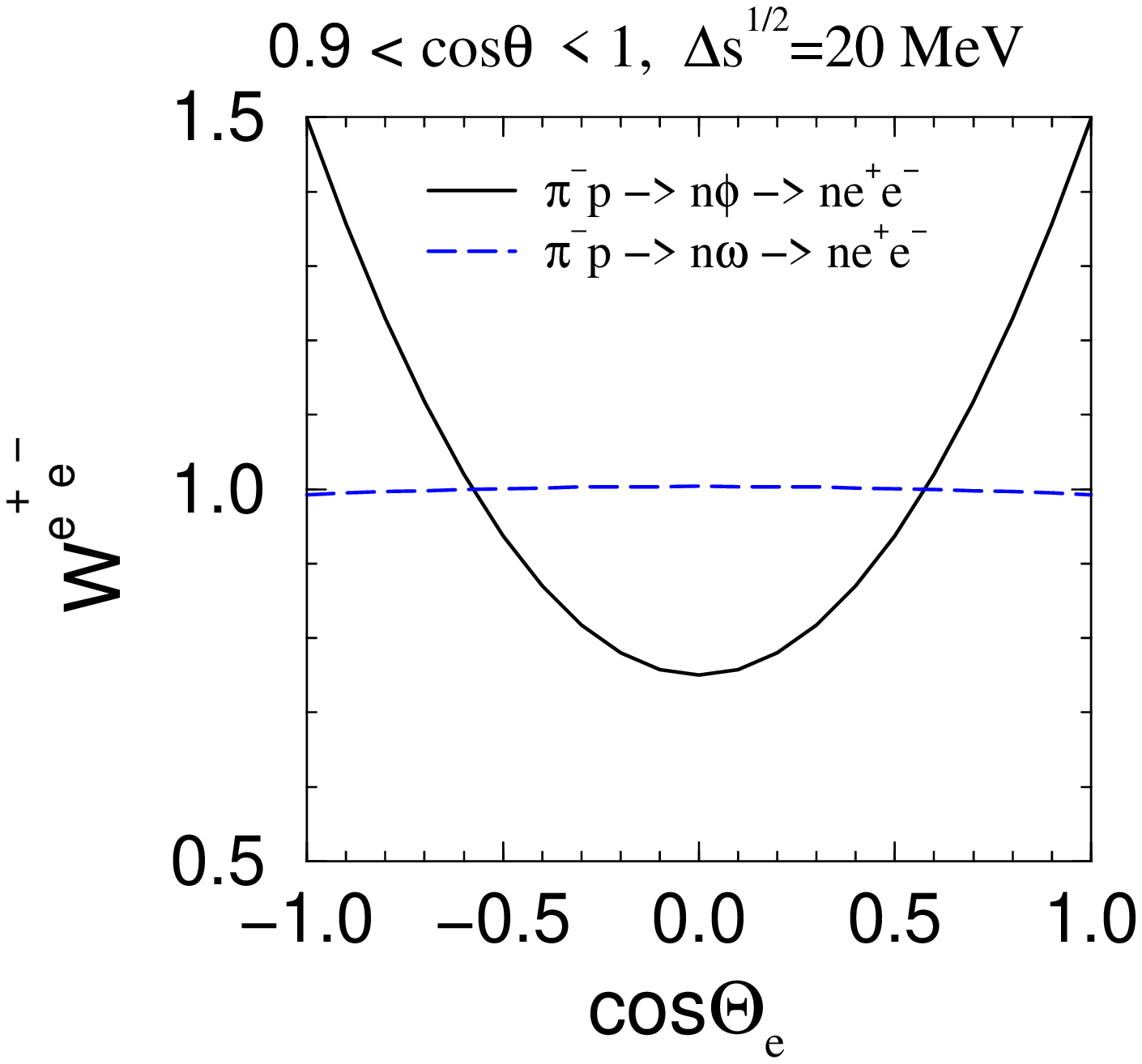, width=7cm}\qquad
 \epsfig{file=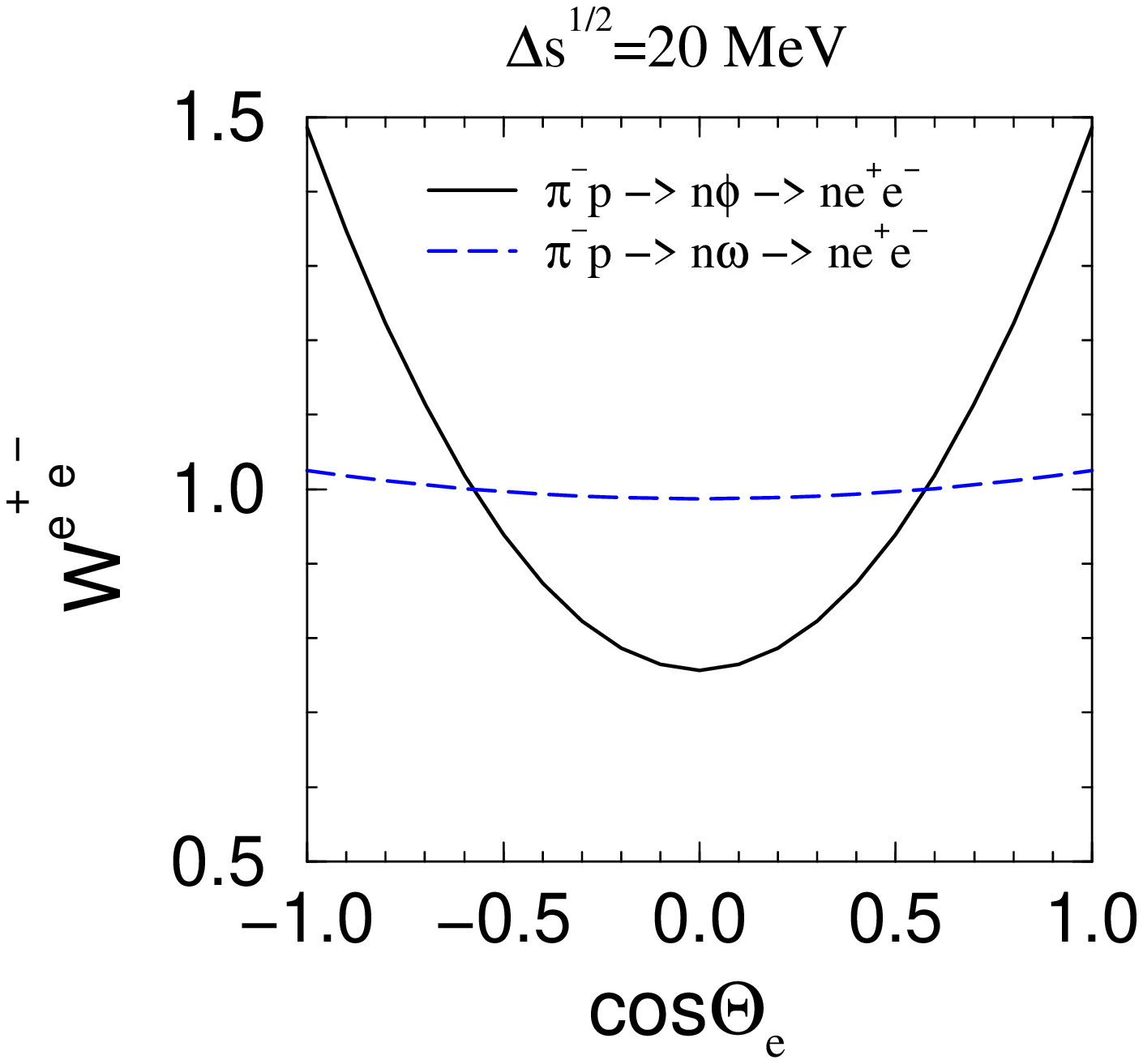, width=7cm}}
\caption{
Electron angular distributions in the reaction
$\pi^- p \to n V\to n e^+e^-$
at $\Delta s^{1/2} = 20$ MeV.
Left panel: the distribution at forward vector meson production angles,
right panel: the distribution averaged over the  all production angles.}
\label{fig:10}
\end{figure}

\begin{figure}
\centering
{\epsfig{file=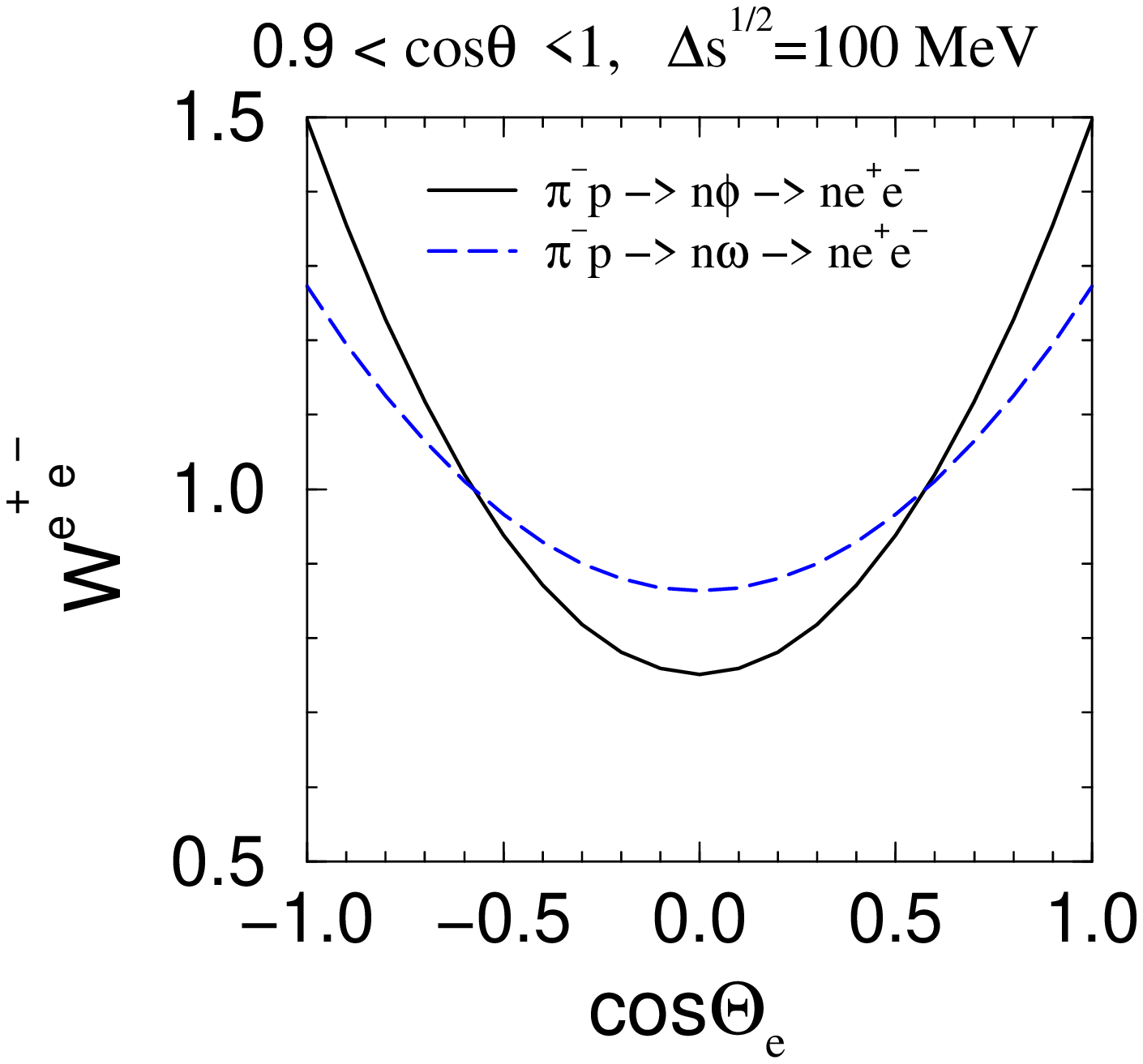, width=7cm}\qquad
 \epsfig{file=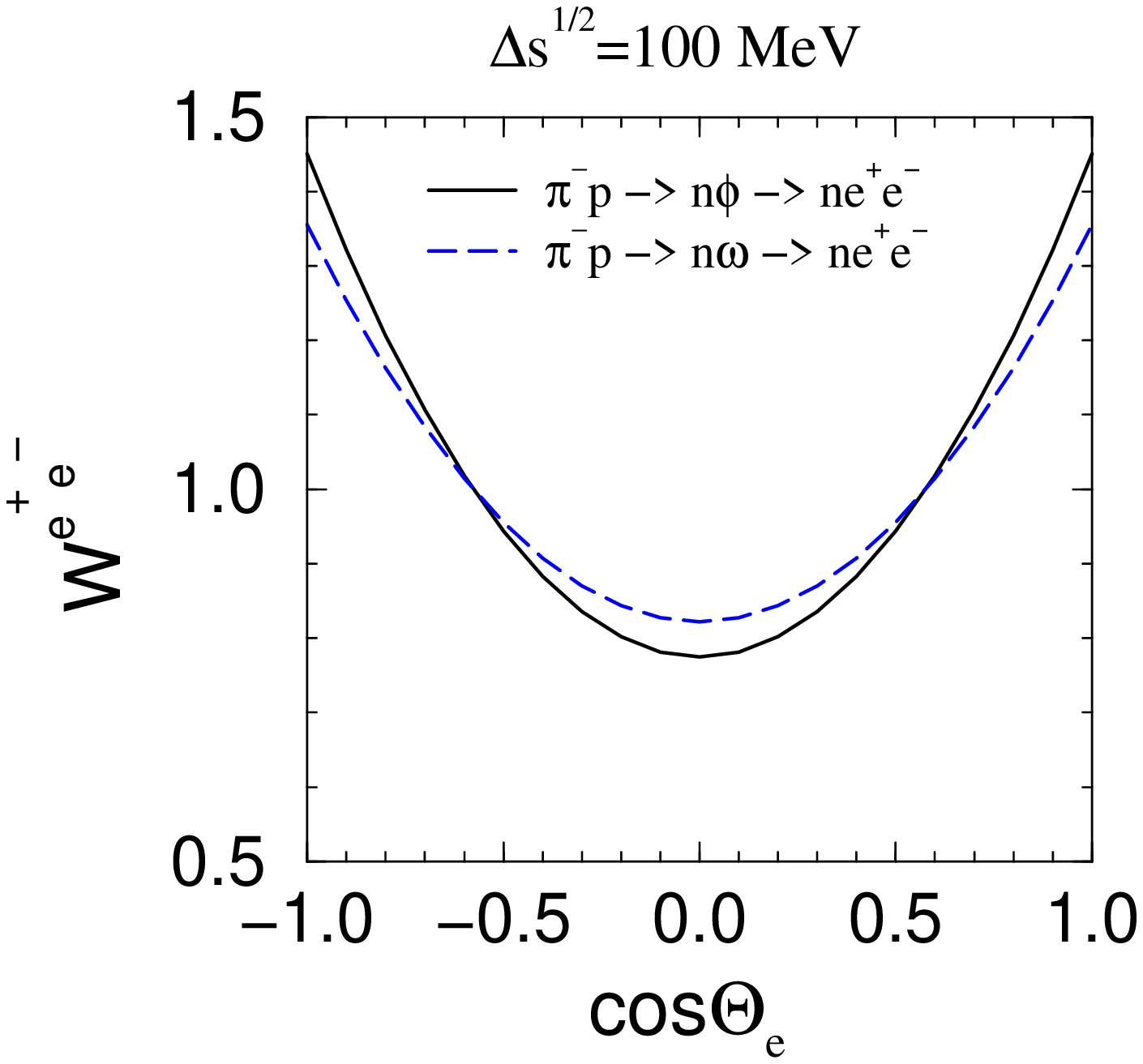, width=7cm}}
\caption{
The same as in Fig.~10 but at $\Delta s^{1/2}=100$ MeV.}
\label{fig:11}
\end{figure}

\end{document}